\documentclass[useAMS,usenatbib]{mn2e}

\usepackage{graphicx}
\usepackage{amsmath}
\usepackage{amssymb}
\usepackage{color}
\usepackage{subfig}
\usepackage[breaklinks,colorlinks,citecolor=blue,linkcolor=red]{hyperref} 
\usepackage[all]{hypcap}

\newcommand\msun{\, \rm M_\odot}
\newcommand\rsun{\, \rm R_\odot}
\newcommand\kms{\, \rm km\,s^{-1}}

\newcommand\mpcub{\, \rm M_\odot\,pc^{-3}}

\newcommand\gpcyr{\, \rm Gpc^{-3}\,yr^{-1}}

\newcommand\mnsc{{M_{\rm NSC}}}
\newcommand\nnsc{{n_{\rm NSC}}}
\newcommand\rhonsc{{\rho_{\rm NSC}}}

\newcommand\be{\begin{equation}}
\newcommand\ee{\end{equation}}

\title[Binary BH population in NSCs through TDEs]{Calibrating the binary black hole population in nuclear star clusters through tidal disruption events}
\author[G. Fragione, R. Perna, \& A. Loeb]{\parbox{\textwidth}{Giacomo Fragione$^{1,2}$\thanks{E-mail: giacomo.fragione@northwestern.edu}, Rosalba Perna$^{3,4}$, Abraham Loeb$^{5}$}\\
\ \\
$^1$Department of Physics \& Astronomy, Northwestern University, Evanston, IL 60202, USA\\
$^2$Center for Interdisciplinary Exploration \& Research in Astrophysics (CIERA), Evanston, IL 60202, USA\\
$^3$Department of Physics and Astronomy, Stony Brook University, Stony Brook, NY 11794-3800, USA\\
$^4$Center for Computational Astrophysics, Flatiron Institute,  New York, NY 10010, USA\\
$^5$Astronomy Department, Harvard University, 60 Garden St., Cambridge, MA 02138, USA}

\begin{document}

\maketitle

\begin{abstract}
As the sensitivity of gravitational wave (GW) instruments improves and new networks start operating, hundreds of merging stellar-mass black holes (SBHs) and intermediate-mass black holes (IMBHs) are expected to be observed in the next few years. The origin and distribution of SBH and IMBH binaries in various dynamical environments is a fundamental scientific question in GW astronomy. In this paper, we discuss ways tidal disruption events (TDEs) may provide a unique electromagnetic window into the assembly and merger of binary SBHs and IMBHs in nuclear star clusters (NSCs). We discuss how the host NSC mass and density and the slope of the black-hole mass function set the orbital properties and the masses of the binaries that undergo a TDE. For typical NSC properties, we predict a TDE rate of $\sim 10^{-6}$--$10^{-7}\ {\rm yr}^{-1}$ per galaxy. The lightcurve of TDEs in NSCs could be interrupted and modulated by the companion black hole on the orbital period of the binary. These should be readily detectable by optical transient surveys such as the Zwicky Transient Facility and LSST.
\end{abstract}

\begin{keywords}
stars: black holes -- galaxies: kinematics and dynamics -- stars: black holes -- stars: kinematics and dynamics -- galaxies: nuclei
\end{keywords}

\section{Introduction}
\label{sect:intro}

Ten binary stellar-mass black holes (SBHs) and one binary neutron star (NS) have been detected through their gravitational wave (GW) emission during the first and second observing runs of LIGO/Virgo\footnote{\url{http://www.ligo.org}} \citep{gwcat2019}. The third run has already revealed the merger of a binary NS of $\sim 3.4\msun$ \citep{gwnsO3}, the most massive ever observed, and the merger of a binary SBH with a low-mass ratio and high spin \citep{gwbhO3}. The merger rate in the local Universe has been estimated to be in the ranges $\sim 9.7$--$101\gpcyr$ and $\sim 250$--$2810\gpcyr$ for binary SBH and NS, respectively. No SBH-NS binaries have been confirmed, with a LIGO/Virgo $90\%$ upper limit of $\sim 610\gpcyr$ on the merger rate \citep{gwcat2019}.

Many astrophysical scenarios have been proposed to explain the mergers observed via GW emission by the LIGO/Virgo collaboration. Possibilities include isolated binary evolution \citep{bel16b,giac2018G,kruc2018}, mergers in star clusters \citep{askar17,baner18,frak18,rod18,sam2018,DiCarlo2019,Perna2019,kr2020ApJS,DiCarlo2020}, mergers in galactic nuclei \citep{olea09,antoper12,fragrish2018,grish18,hamil2019,rass2019}, mergers in gaseous disks \citep{bart17,Tagawa2019}, and Lidov-Kozai (LK) mergers in isolated triple and quadruple systems \citep{ant17,sil17,fragk2019,hamers2019,liu2019}.

Merger of black holes of $\sim 10^2\msun$--$10^5\msun$, in the regime of intermediate-mass black holes (IMBHs), could also be detected by GW instruments. At design sensitivity, LIGO/Virgo, the Einstein Telescope\footnote{\url{http://www.et-gw.eu}} (ET), DECIGO\footnote{\url{https://decigo.jp/index_E.html}} and LISA\footnote{\url{https://lisa.nasa.gov}}, will be able to detect GW sources from merging IMBHs of masses up to $\sim 100-1000\msun$, $\sim 10^3-10^4\msun$ and $\gtrsim 10^4\msun$, respectively \citep[e.g.,][]{seoane2007,fraginkoc18,bello2019}. Using the non-detection of massive binaries in the first two observational runs, the LIGO/Virgo collaboration placed upper limits on merging IMBHs of the order of $\sim 0.1$--$1\gpcyr$ \citep{ligov2019}.

Mergers of binaries where one or both  components are IMBHs can occur in a variety of environments. Unlike the SBH case, binaries hosting an IMBH cannot be produced by the collapse of stars in a canonical binary evolution, except for the case of Pop III stars \citep{madau2001,bromm2004,fryer2001,bromm2013,loebfur2013}. Therefore, dynamical scenarios are generally favored, representing the ideal place where an IMBH can find a SBH/NS/IMBH companion to merge with. These environments include nuclear star clusters \citep{mast14,frle2018,anto2019}, globular clusters \citep{mandel2008,fraginkoc18,rassfk2019,kretal2020}, and active galactic nuclei (AGN) accretion disks \citep{McKernan+2012,McKernan+2014}.

As the sensitivity of current GW detectors improves and new networks start operating, hundreds of merging SBHs are expected to be detected over the next few years \citep{gwcat2019}. While low-mass IMBHs could already be observed by LIGO/Virgo \citep{ligov2019}, the majority of IMBHs merging in binary systems are expected to be detected with the upcoming ET, DECIGO, and LISA missions \citep{miller2009}. Therefore, the origin and distribution of SBH and IMBH binaries in dynamical environments is a key scientific question that will be addressed with future GW data.

In this paper we discuss the electromagnetic window, provided by tidal disruption events (TDEs) of stars, into the assembly and merger history of binary SBHs and IMBHs\footnote{In the following, we use BH to refer to either SBH or IMBH without distinction.} in nuclear star clusters (NSCs). In this environment the lightcurve of the TDEs can be interrupted and modulated by the companion BH on the orbital period of the binary \citep[e.g.,][]{liuli2009,cough2017,fraglpk2019}, which can then be used to probe the binary SBH and IMBH orbital period distribution \citep{samsing2019}. We compute the rates of TDEs for various NSC masses and densities, and different distributions of the SBH and IMBH mass function. We also discuss the typical electromagnetic signal expected as a result of these events and how it can be used to detect binary BHs.

The paper is organized as follows. In Section~\ref{sec:bhbin}, we describe the various channels that lead to the formation of binary BHs. In Section~\ref{sec:tdes}, we show how TDEs can be used to infer the characteristics of binary BHs in galactic nuclei, and compute the relative rates for a population of SBHs and IMBHs. In Section~\ref{sec:electr}, we discuss the electromagnetic signatures of these events. Finally, in Section~\ref{sec:conc}, we discuss the implications of our findings and draw our conclusions.

\section{Black hole binaries in nuclear star clusters}
\label{sec:bhbin}

Here we summarize the main mechanisms that can lead to the formation of hard SBH-SBH, IMBH-SBH, and IMBH-IMBH binaries in a given NSC. We label the mass and the density of the NSC $\mnsc$ and $\rhonsc$, respectively. We define the cluster velocity dispersion $\sigma=v_{\rm esc}/(2\sqrt{3})$, where $v_{\rm esc}$ is the cluster escape speed \citep{georg2009},
\begin{equation}
v_{\rm esc}=40\kms \left(\frac{\mnsc}{10^5\msun}\right)^{1/3} \left(\frac{\rhonsc}{10^5\msun\mathrm{pc}^{-3}}\right)^{1/6}\,.
\label{eqn:mvesc}
\end{equation}

Binary BHs can be formed through the interaction of three objects in the core of a NSC. Interactions come in two flavours: encounters between three single objects and encounters between a single and a binary \citep[e.g.,][]{antoras2016,fragsilk2020}. In the first case, BH binaries can be assembled through three-body processes in which a binary is formed with the help of a third BH, which carries away the excess energy needed to bind the pair \citep[e.g.,][]{lee1995}. In the second case, binary BHs can also form through exchange interactions, mediated initially by primordial stellar binaries. When a BH gets within a couple of semimajor axes lengths of a binary, it will tend to break the binary and to acquire a companion. After BH binaries are formed, they dominate the dynamics inside the cluster core \citep[e.g.][]{mill2009}.

Another channel to form binary BHs in a NSC is gravitational  bremsstrahlung. In this scenario, two single BHs pass sufficiently close to each other to dissipate enough energy via GW radiation to remain bound \citep[e.g.,][]{Turner1977}.  However, binary BHs formed through GW captures are typically very eccentric and merge within a few minutes up to a few years (depending on the BH masses) after formation \citep{olea09,rasskoc2019,frag2020}.

Star clusters are promising environments for forming binary BHs \citep{askar17,fragkoc2018,rod18}. For what concerns IMBHs, a number of studies have showed that the most massive stars may segregate and merge in the core of the cluster, forming a massive growing object that could collapse to form an IMBH \citep{por02,gurk2004,gie15}. Binaries containing the IMBH can later be formed through three-body interactions \citep{mandel2008,fragbrom2019}. Star clusters born in the innermost galactic regions could have dynamical friction timescales small enough to efficiently inspiral into the centres of galaxies \citep[e.g.,][]{tremaine1975,capuzz2008,gne14}. Any binary BH formed within these clusters could therefore be delivered to the innermost regions of the host galaxy \citep[e.g.,][]{gurk2005,mast14,agu18,fraginkoc18}.

Binary BHs could also form efficiently in the gaseous disks of active galactic nuclei \citep[AGN;][]{McKernan+2012,McKernan+2014}. If migration traps are present in the gaseous disk surrounding a supermassive BH, differential gas torques exerted on the orbiting SBHs and IMBHs will cause them to migrate towards a migration trap \citep{secunda18}. Turbulence in the gaseous disk can knock orbiting SBHs out of resonance, allowing them to drift close to the trap and experience a close interaction with other BHs. Since the interactions are dissipative due to the gas, it could be possible to form SBH-SBH, IMBH-SBH, and IMBH-IMBH binaries. Some of these objects could also merge repeatedly, thus forming massive BHs \citep{mcker2019,yang2019}.

We consider stars undergoing TDEs onto hard binaries that are formed through the interaction of initially unbound BHs. Hard binaries can be defined based on the softness parameter \citep{Heggie1975},
\begin{equation}
\eta=\frac{Gm_{\rm BH,1} m_{\rm BH,2}}{2 a\langle m\rangle \sigma^2}\,,
\label{eqn:etahard}
\end{equation}
where $m_{\rm BH,1}$ is the mass of the primary BH, $m_{\rm BH,2}$ ($m_{\rm BH,2}<m_{\rm BH,1}$) is the mass of the secondary BH, $a$ and $e$ the semi-major axis and eccentricity of the binary, respectively, and $\langle m \rangle$ and $\sigma$ are the average star mass and the cluster velocity dispersion, respectively. Binaries with $\eta\ll 1$ are called soft binaries and will become even softer on average, until they are disrupted by interactions with other stars and compact objects; on the other hand, binaries with $\eta\gg 1$ are referred to as hard binaries and tend to become even harder \citep{Heggie1975}. As a result of their interactions with other stars and compact objects, binaries formed at the hardening semi-major axis \citep{quin1996},
\begin{equation}
a_{\rm hard}=2.5\ \mathrm{AU}\left(\frac{m_{\rm BH,2}}{10\msun}\right)\left(\frac{\sigma}{30\kms}\right)^{-2}\,,
\label{eqn:ahard}
\end{equation}
tend to shrink at a constant rate \citep{quin1996}, eventually down to a separation,
\begin{eqnarray}
a_{\rm GW}&=&0.05\ \mathrm{AU}\left(\frac{m_{\rm BH,1}+m_{\rm BH,2}}{20\msun}\right)^{3/5}\left(\frac{10^6\msun\ \mathrm{pc}^{-3}}{\rhonsc}\right)^{1/5}\times \nonumber\\
&\times& \left(\frac{\sigma}{30\kms}\right)^{1/5}\left(\frac{q}{(1+q)^2}\right)^{1/5}\,,
\label{eqn:agw}
\end{eqnarray}
where GWs takes over and hence the binary merges. Note that during one of the interactions that makes the binary shrink, the binary itself can receive a dynamical kick such that it is ejected from the host NSC \citep{antoras2016}.

\section{Tidal disruption events in binary black holes}
\label{sec:tdes}

More than $\sim 60\%$ of early- and late-type galaxies show observational evidence indicating that the nuclear regions of galaxies are occupied by a NSC and/or a central supermassive black hole. NSCs are present mainly in faint galaxies, while supermassive black holes are common in galaxies with masses $\gtrsim 10^{10}\msun$ \citep[e.g.,][]{capuzzo2017}. The most complete catalog of NSC can be found in the sample of \citet{georg2016}. This sample of NSCs comprises the systems in spheroid-dominated galaxies \citep{cote2006,turner2012} and disc-dominated galaxies \citep{georg2009,geor2014}. The masses and densities of the known NSCs are $10^4\msun\lesssim \mnsc \lesssim 10^8\msun$ and $10^4\mpcub\lesssim \rhonsc \lesssim 10^8\mpcub$, respectively \citep[see e.g. Fig. 1 in][]{fragsilk2020}.

\begin{figure*} 
\centering
\includegraphics[scale=0.55]{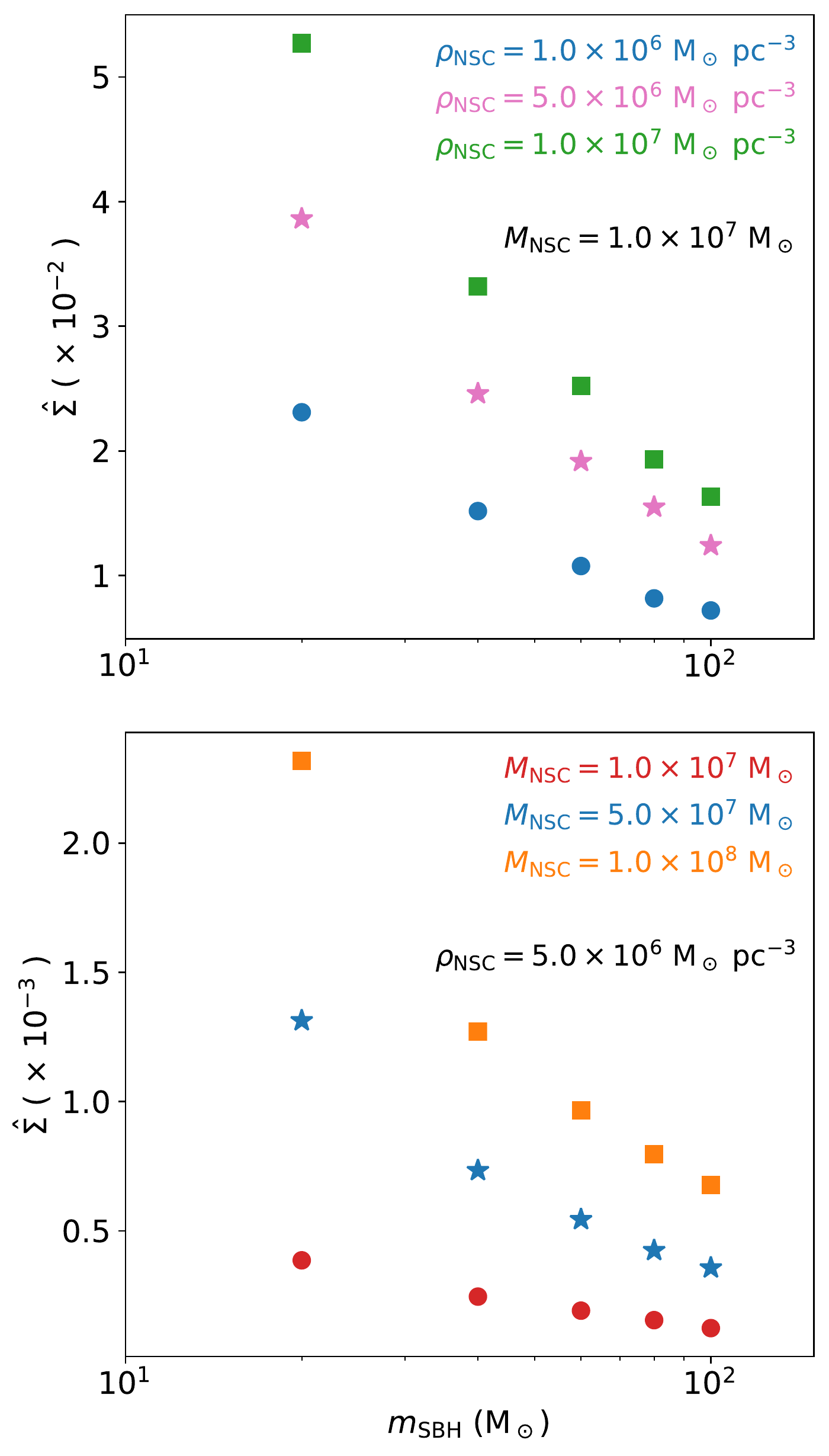}
\hspace{0.5cm}
\includegraphics[scale=0.55]{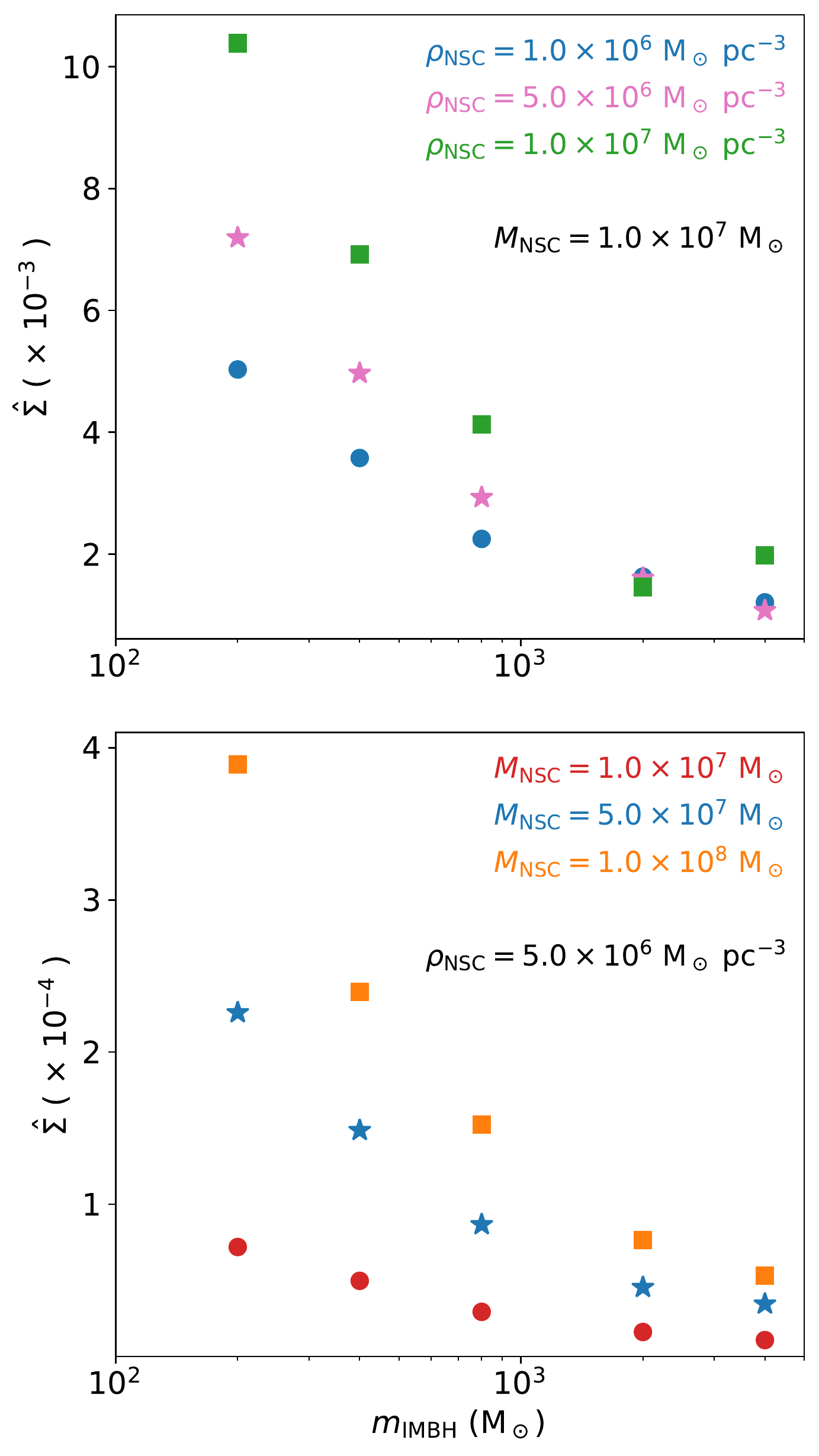}
\caption{Left: branching ratio ($\hat{\Sigma}$) as a function of the SBH mass ($m_{\rm SBH}$) for equal mass binaries at the hardening radius. Right: branching ratio as a function of the IMBH mass ($m_{\rm IMBH}$) for equal mass binaries at the hardening radius. Top panel: $\mnsc=10^7\msun$ and different values of $\rhonsc$; bottom panel: $\rhonsc=5\times 10^6\mpcub$ and different values of $\mnsc$.}
\label{fig:bhmx}
\end{figure*}

\begin{figure} 
\centering
\includegraphics[scale=0.575]{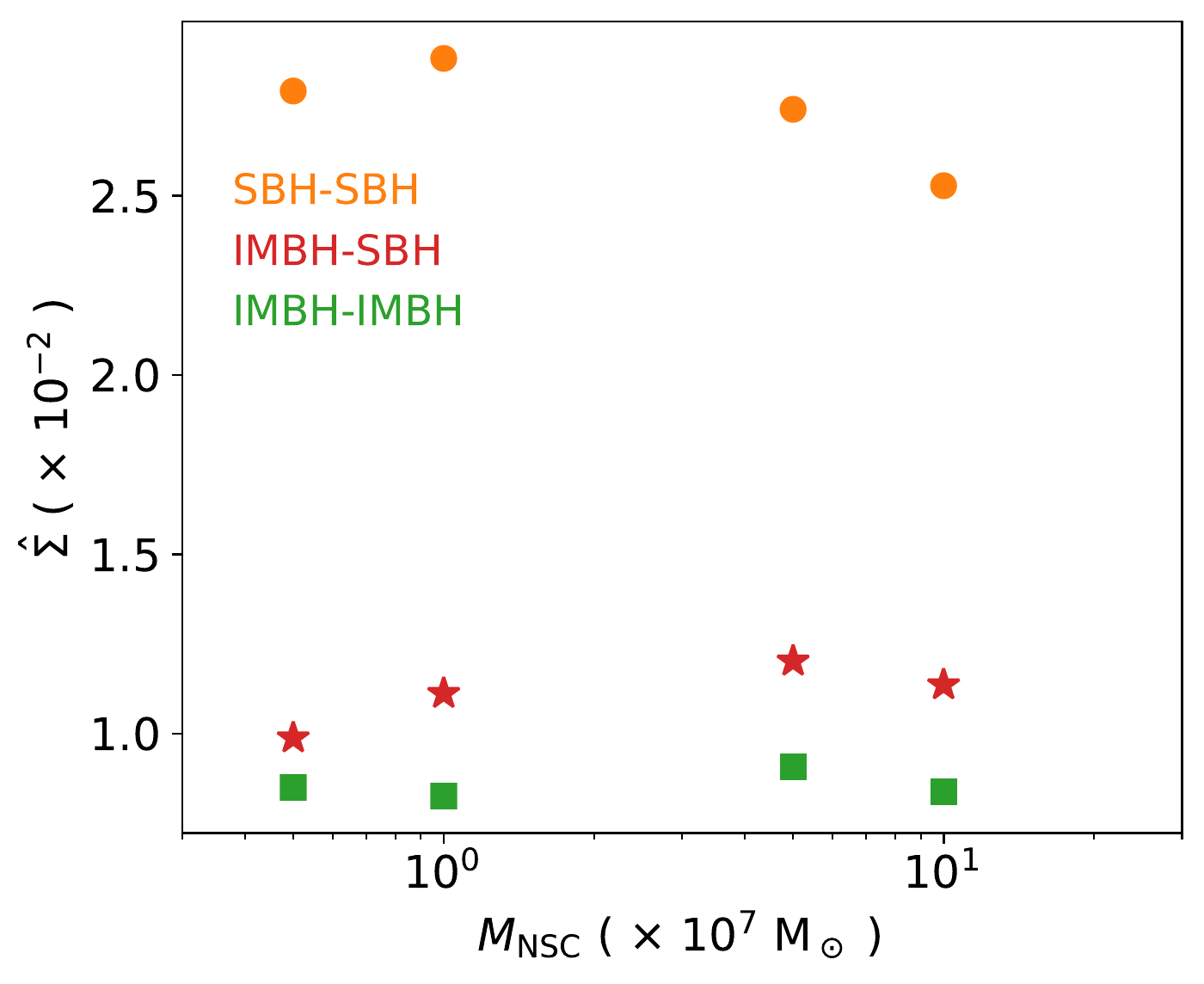}
\includegraphics[scale=0.575]{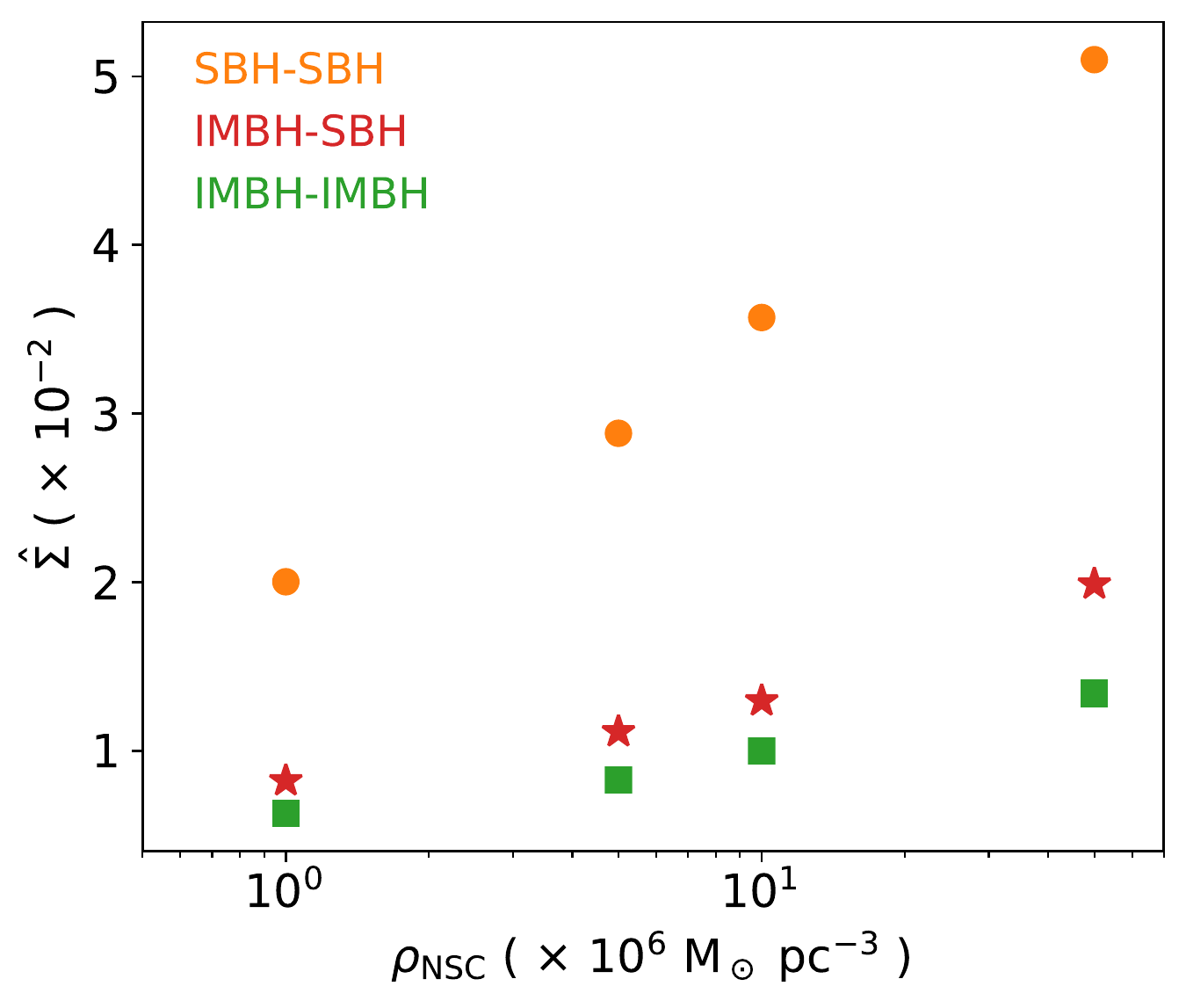}
\includegraphics[scale=0.575]{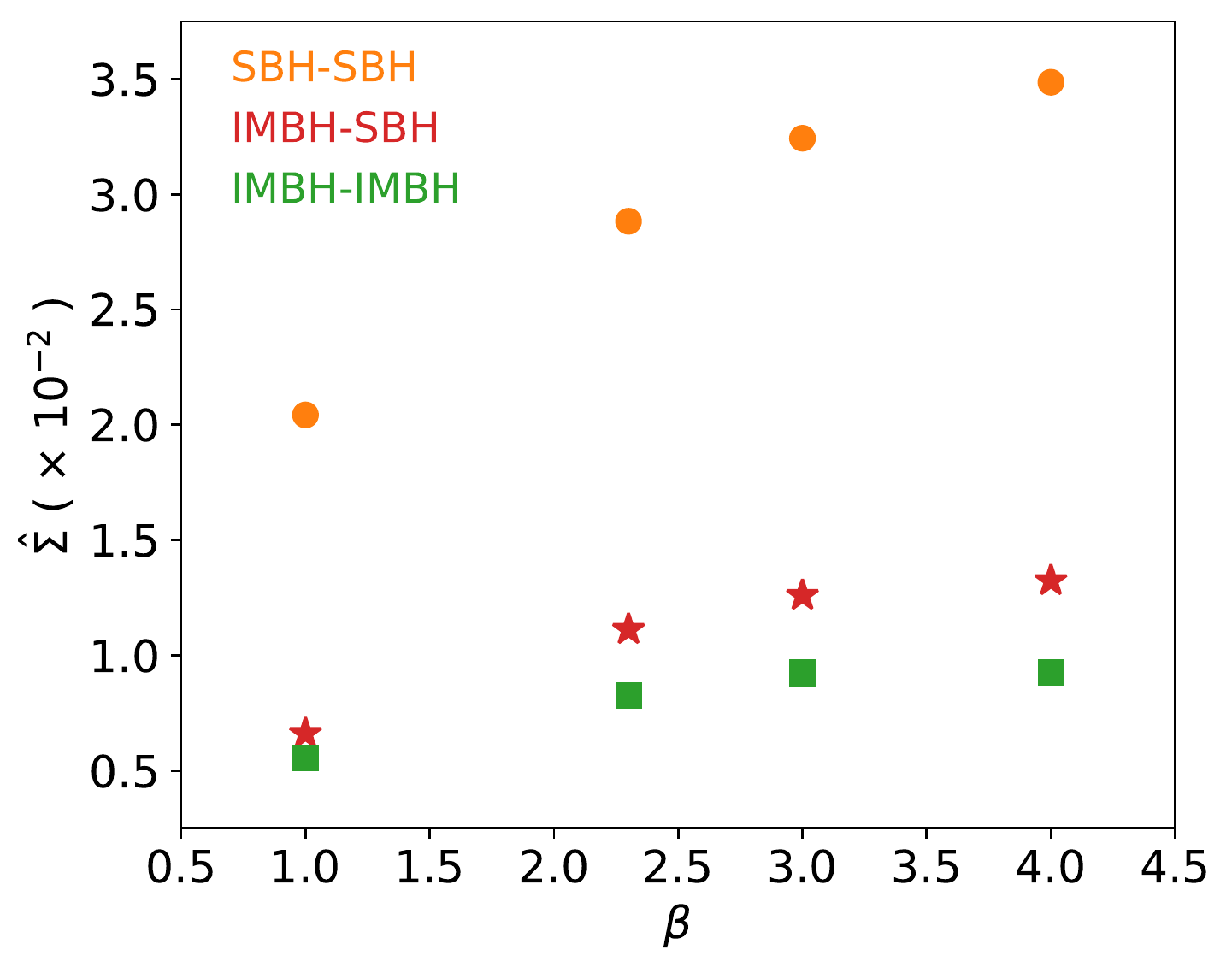}
\caption{Top panel: branching ratio as a function of the NSC mass ($\rhonsc=5\times 10^6\mpcub$, $\beta=2.3$) for SBH-SBH (orange), IMBH-SBH (red), and IMBH-IMBH (green). Central panel: branching ratio as a function of the NSC density ($\mnsc=10^7\msun$, $\beta=2.3$) for SBH-SBH (orange), IMBH-SBH (red), and IMBH-IMBH (green). Bottom panel: branching ratio as a function of the the slope of the BH mass function ($\mnsc=10^7\msun$, $\rhonsc=5\times 10^6\mpcub$) for SBH-SBH (orange), IMBH-SBH (red), and IMBH-IMBH (green).}
\label{fig:bhmassdensbeta}
\end{figure}

We consider the fate of a star (assumed to have a solar mass, $m_*=1\msun$, and radius $R_*=1\rsun$) that interacts with a binary SBH-SBH, IMBH-SBH, or IMBH-IMBH during its lifetime in NSCs of various masses and densities. Table~\ref{tab:models} summarizes the properties of the host NSC we consider in this work:

\begin{itemize}

\item The NSC mass and density are varied in the range $0.5\le \mnsc/10^7 \msun \le 10$ and $1\le \rhonsc/10^6 \mpcub\le 50$, respectively.

\item BH masses are sampled from a BH mass function, assumed for simplicity to follow a distribution
\begin{equation}
f(m_{\rm BH})\propto m_{\rm BH}^{-\beta}\,,
\end{equation}
where $1.0\le\beta\le 4$. SBH masses are drawn in the interval $[5\msun,100\msun]$, while IMBH masses in the interval $[10^2\msun,10^4\msun]$.

\item The initial BH binaries are assumed to be on a circular orbit of initial semi-major axis chosen either to be $a=a_{\rm hard}$ (Mod1 and Mod2) or in the range $a_{\rm GW}<a<a_{\rm hard}$ (Mod3, Mod4, and Mod5).

\item In the scattering experiments, the relative velocity of the binary and of the incoming star are fixed to the velocity dispersion $\sigma$ of the host NSC, assumed to be roughly constant within the region of interest \citep{pechetti2020}. The impact parameter is drawn from a distribution
\begin{equation}
f(b)=\frac{1}{2}\frac{b}{b_{\rm max}^2}\,,
\end{equation}
where $b_{\rm max}$ is the maximum impact parameter of the scattering experiment defined by
\begin{equation}
b_{\rm max}=p_{\rm max}\sqrt{1+\frac{2G M_T}{p_{\rm max} \sigma^2}}\ .
\label{eqn:bmax}
\end{equation}
In the above equation, $M_T$ is the total mass (binary+star) of the system and $p_{\rm max}$ is the maximum pericentre distance of the encounter. The initial separation of the binary and incoming star is chosen to be the distance at which the tidal perturbation on both systems has a fractional amplitude $\delta=F_{\rm tid}/F_{\rm rel}=10^{-5}$. Here, $F_{\rm rel}$ and $F_{\rm tid}$ are the relative forces between the components of the binary and the initial tidal force between the binary and the star, respectively \citep{fregeau2004,antogn16}.

\item Two angles describe the relative orientations and phases of the encounter, given the plane of motion of the center of mass of the BH binary and of the star. The relative inclination of the orbital plane of the binary constitutes the first angle. The initial phases of the BHs in the binary add an additional angle. For all the scattering experiments, these angles are chosen randomly.

\end{itemize}

To run our scattering experiments, we use the \texttt{FEWBODY} numerical toolkit \citep{fregeau2004}, which classifies the results of the scattering events into a set of independently bound hierarchies and considers a run completed when their relative energy is positive and the tidal perturbation on each outcome system is smaller than $\delta$.

As discussed in \citet{chen2009} and \citet{chen2011} in the context of supermassive black hole binaries, close resonant encounters and secular effects of the stars that are trapped within the binary orbit can efficiently produce TDEs. To quantify the probability of TDEs, we define the cross section
\begin{equation}
\Sigma_{\rm TDE}=\pi b_{\rm max}^2\hat{\Sigma}_{\rm TDE}\,,
\end{equation}
where $b_{\rm max}$ is the maximum impact parameter and
\begin{equation}
\hat{\Sigma}_{\rm TDE}=\frac{N_{\rm TDE}}{N_{\rm tot}}
\end{equation}
is the branching ratio. Here, $N_{\rm TDE}$ is the number of scatterings that ends in a TDE, which is determined by the condition that the star passes within the tidal disruption radius
\begin{equation}
R_{\rm T}=R_*\left(\frac{m_{\rm BH,i}}{m_*}\right)^{1/3}\,,\ \ \ i=1,2
\label{eq:Rt}
\end{equation}
of one of the BHs ($i=1,2$) in the binary, and $N_{\rm tot}$ is the total number of scattering experiments. In our simulations, we fix $b_{\rm max}=2a_{\rm hard}$, since the stars that result in a TDE are the ones that can be effectively deflected and trapped in the binary BH orbit. For each model in Table~\ref{tab:models}, we run $10^5$ scattering experiments.

\subsection{Dependence on the maximum black hole mass}
\label{subsect:bhmax}

In Mod1 and Mod2 (see Table~\ref{tab:models}), we consider scattering experiments of stars interacting with equal-mass binaries. We consider both SBH-SBH and IMBH-IMBH binaries. In order to estimate the role of the maximum BH mass, we fix the binary semi-major axis to the hardening semi-major axis $a_{\rm hard}$ (Eq.~\ref{eqn:ahard}).

\begin{table}
\caption{Model parameters: name, mass of the NSC ($
\mnsc$), density of the NSC ($\rhonsc$), slope of the black hole mass function ($\beta$).}
\centering
\begin{tabular}{lccccc}
\hline
Name & $\mnsc$ ($10^7 \msun$) & $\rhonsc$ ($10^6 \mpcub$) & $\beta$ \\
\hline\hline
Mod1  &   $1$           &   $1$-$10$  &   - \\  
Mod2  &   $1$--$10$     &   $5$       &   - \\  
Mod3  &   $0.5$--$10$   &   $1$       &   $2.3$ \\  
Mod4  &   $0.5$         &   $1$--$50$ &   $2.3$ \\  
Mod5  &   $0.5$         &   $1$       &   $1.0$-$4.0$ \\
\hline
\end{tabular}
\label{tab:models}
\end{table}

We plot in Figure~\ref{fig:bhmx} the branching ratio $\hat{\Sigma}$ as a function of the SBH mass $m_{\rm SBH}$ (left panel) and as a function of the IMBH mass $m_{\rm IMBH}$ (right panel) for equal mass binaries. In the top panel we show the branching ratio for TDEs in the case the host NSC has mass $\mnsc=10^7\msun$ and different densities, while in the bottom panel we show the branching ratio for TDEs in the case the host NSC has density $\rhonsc=5\times 10^6\mpcub$ and different masses. In both cases, we find that $\hat{\Sigma}$ decreases for larger BH masses. Moreover, the branching ratio becomes larger for larger NSC masses and densities. These trends can be understood in terms of the scale of the hardening radius,
\begin{equation}
a_{\rm hard}\propto \frac{m_{\rm BH}}{\sigma^2}\propto \frac{m_{\rm BH}}{M_{\rm NSC}^{2/3}\rho_{\rm NSC}^{1/3}}\,.
\end{equation}
For fixed NSC mass and density, larger BH masses correspond to larger hardening radii; on the other hand, larger NSC masses and densities lead to larger hardening radii, at a fixed BH mass. While wider binaries can intercept a larger flux of incoming stars, the stars that lead to a TDE are the ones that, during resonant encounters, have the chance to pass within the tidal radius (see Eq.~\ref{eq:Rt}) of one of the two BHs ($m_{\rm BH}$) and be tidally disrupted. As a result, the branching ratio is smaller for wider binaries, since it will be less probable for a star to pass close enough ($\lesssim R_{\rm T}$) to one of the BHs during resonant encounters.

\subsection{Dependence on NSC mass and density and slope of black hole mass function}
\label{subsect:massdensnsc}

In Mod3, Mod4, and Mod5, we consider a population of SBH-SBH, IMBH-SBH, IMBH-IMBH binaries to study the effect of the NSC mass, NSC density, and slope of the mass function. In these models, we consider binaries that are hard enough not to be ionized by encounters with other stars and compact objects ($a<a_{\rm hard}$), but wide enough such that their dynamics is not dominated by GW emission ($a>a_{\rm GW}$). To this end, we sample the semi-major axis of the interacting binary BHs from a log-uniform distribution in the interval $(a_{\rm GW},a_{\rm hard})$. BH masses are sampled from a negative power law with exponent $\beta$.

We present the results of our scattering experiments in Figure~\ref{fig:bhmassdensbeta}, for SBH-SBH (orange), IMBH-SBH (red), and IMBH-IMBH (green) binaries. In the top panel we show $\hat{\Sigma}$ as a function of the NSC mass ($\rhonsc=5\times 10^6\mpcub$, $\beta=2.3$); in the central panel, we plot the branching ratio as a function of the NSC density ($\mnsc=10^7\msun$, $\beta=2.3$); in the bottom panel, we show $\hat{\Sigma}$ as a function of the the slope of the BH mass function ($\mnsc=10^7\msun$, $\rhonsc=5\times 10^6\mpcub$). The general trend with $a_{\rm hard}$ discussed previously is in part smeared out by the fact that we sample the binary semi-major axis in the range $(a_{\rm GW},a_{\rm hard})$, rather than having it fixed to $a_{\rm hard}$. Moreover, BHs do not have a fixed mass, as in Mod1 and Mod2 in equal-mass binaries, but their masses are sampled from a distribution. Additionally, we find that SBH-SBH binaries produce the largest branching ratio, while IMBH-IMBH lead to the lowest one. This can again be ascribed to the fact that hard binaries of two SBHs have typically smaller semi-major axes than binaries comprised of two IMBHs. As a consequence, stars that undergo a resonant encounter with an SBH-SBH binary are more likely to pass within the tidal radius of one of the two BHs and get tidally disrupted. Branching ratios of binary IMBH-SBHs are in between the SBH-SBH and IMBH-SBH cases.

\begin{figure} 
\centering
\includegraphics[scale=0.55]{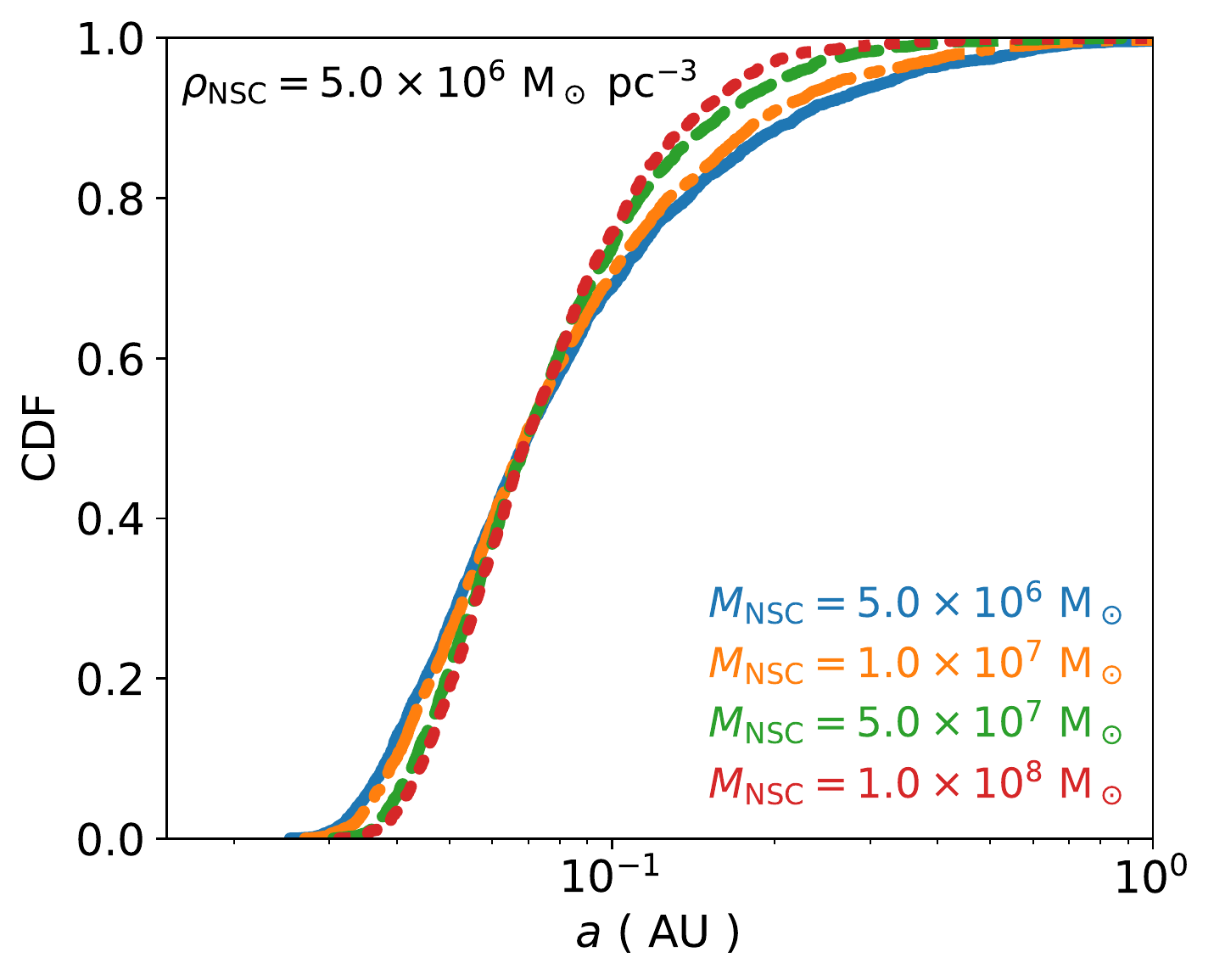}
\includegraphics[scale=0.55]{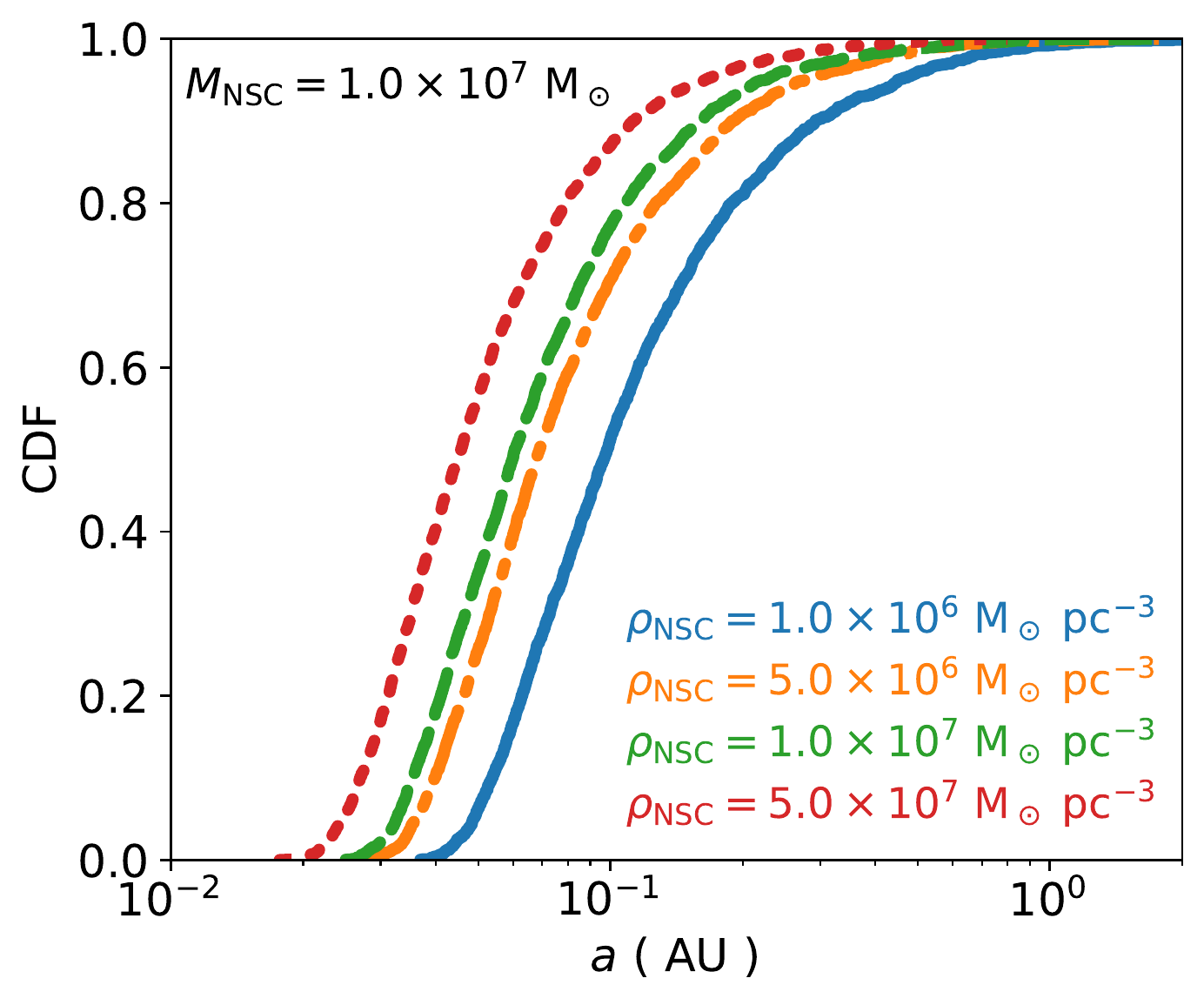}
\caption{Distribution of semi-major axis ($a$) of the SBH-SBH binaries that undergo a TDE. Top panel: $\rhonsc=5\times 10^6\mpcub$ and different values of the NSC mass; bottom panel: $\mnsc=10^7\msun$ and different values of the NSC density. The slope of the SBH mass function is fixed to $\beta=2.3$.}
\label{fig:bhparall1}
\end{figure}

\begin{figure} 
\centering
\includegraphics[scale=0.55]{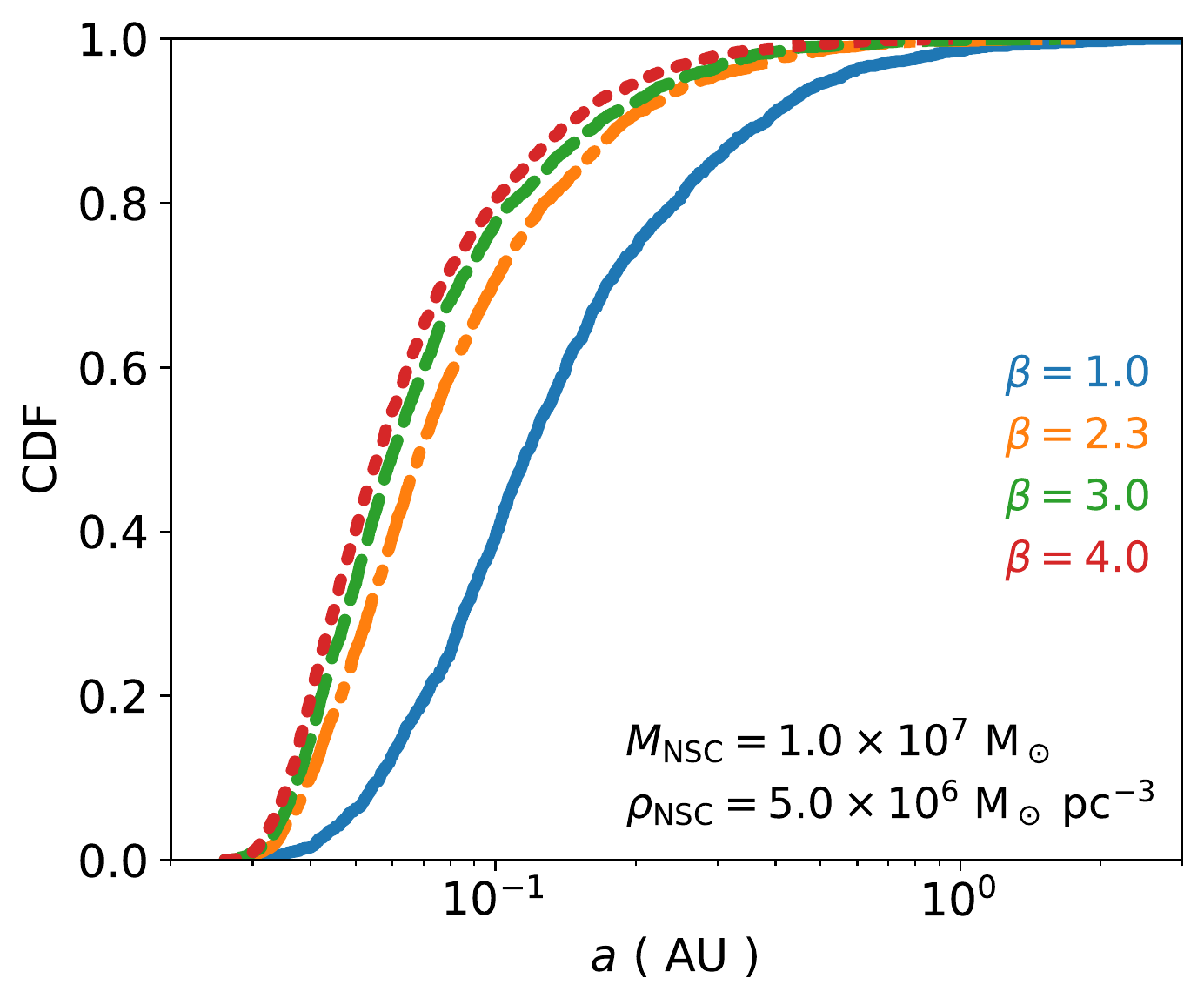}
\includegraphics[scale=0.55]{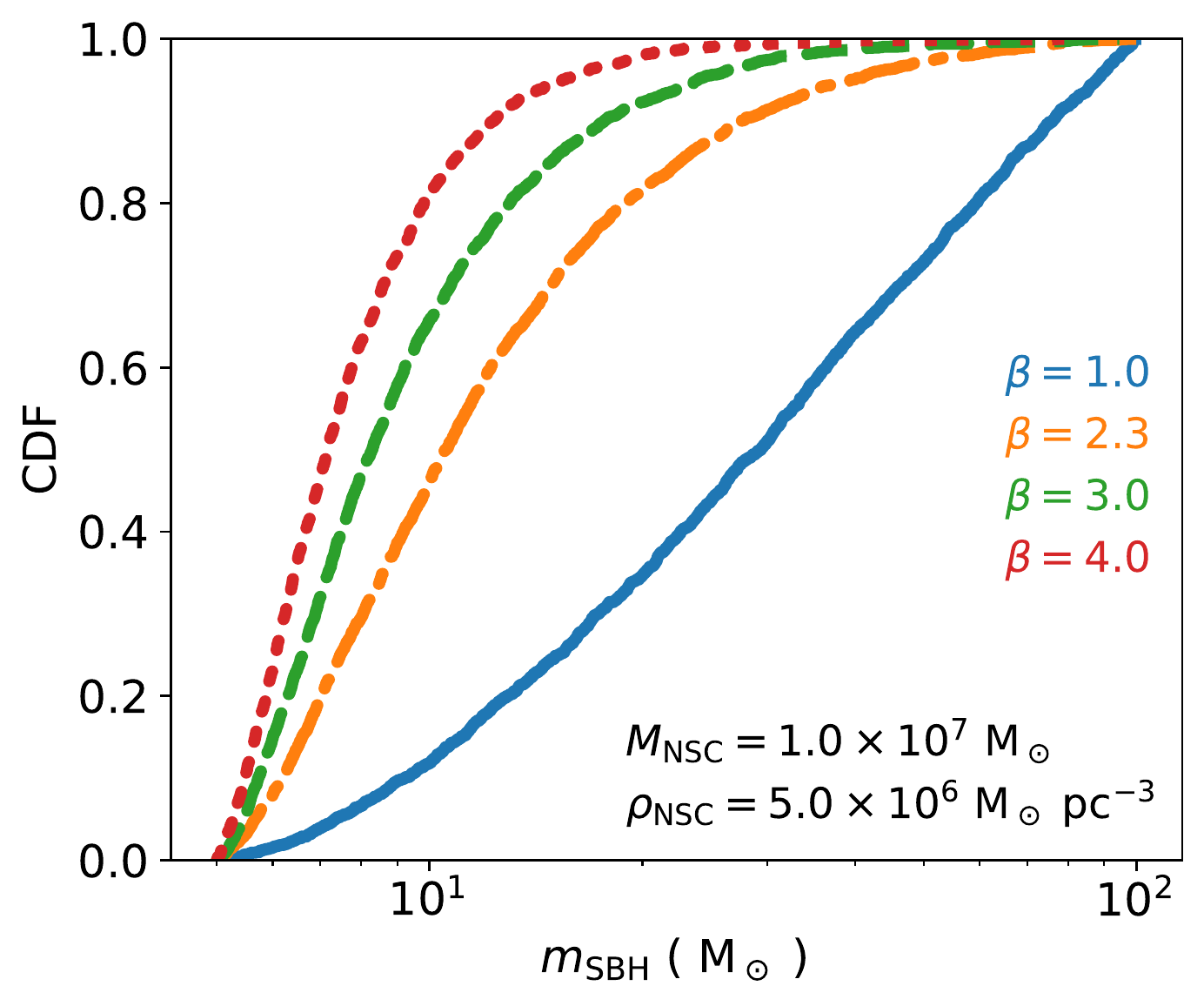}
\caption{Distribution of semi-major axis ($a$) of the SBH-SBH binaries (top) and masses ($m_{\rm SBH}$) of the SBHs (bottom) that undergo a TDE, for different values of the slope of the SBH mass function. The mass and density of the NSC are fixed to $\mnsc=10^6\msun$ and $\rhonsc=5\times 10^6\mpcub$, respectively.}
\label{fig:bhparall2}
\end{figure}

\subsection{Orbital properties of binaries that exhibit tidal disruption events}
\label{subsect:orbitalprop}

The semi-major axis of the binaries that undergo a TDE depends on the host NSC mass and density, and on the slope of the BH mass function $\beta$.

We illustrate in Figure~\ref{fig:bhparall1} the distribution of semi-major axes of the SBH-SBH binaries that undergo a TDE. In the top panel, we show the case for $\rhonsc=5\times 10^6\mpcub$ and different values of the NSC mass (Mod3), while in the bottom panel, we plot the case for $\mnsc=10^7\msun$ and different values of the NSC density (Mod4). The slope of the SBH mass function is fixed at $\beta=2.3$. We find that the NSC mass has only a modest effect in shaping the semi-major axis of the binaries that undergo a TDE, unlike the host NSC density. The reason for this trend can be explained considering that binaries have orbital semi-major axis in the range $(a_{\rm GW},a_{\rm hard})$, where
\begin{equation}
a_{\rm GW}\propto \frac{M_{\rm NSC}^{1/15}}{\rho_{\rm NSC}^{1/6}}\,.
\end{equation}
Larger densities correspond to smaller values of $a_{\rm GW}$, thus a wider range of possible binary semi-major axes, while the NSC mass does not have a significant effect.

Figure~\ref{fig:bhparall2} shows the distribution of semi-major axes ($a$) of the SBH-SBH binaries (top) and masses ($m_{\rm SBH}$) of the SBHs (bottom) that undergo a TDE, for different values of the slope of the SBH mass function. The mass and density of the NSC are fixed to $\mnsc=10^6\msun$ and $\rhonsc=5\times 10^6\mpcub$, respectively. We find that the typical semi-major axis of the binaries that lead to a TDE are larger for smaller values of $\beta$. Shallower BH mass functions (smaller $\beta$'s) produce on average more massive BHs, which can form hard binaries at larger separations ($a_{\rm hard}\propto m_{\rm SBH}$). The slope of the mass function also obviously affect the mass of the BH that undergoes TDE, producing less massive BHs for steeper mass functions. The NSC mass and density do not affect the SBH mass distribution.

The same trends discussed for SBH-SBH binaries hold for IMBH-SBH and IMBH-IMBH binaries, with the only difference that their typical semi-major axes are larger, being more massive than SBH-SBH binaries.

\subsection{Rates}
\label{subsect:rates}

We now turn to the astrophysical rates of stellar TDEs
onto binary BHs expected from NSCs. The rate per galaxy depends both on the mass and density of the host NSC and the absolute number of binary BHs ($N_{\rm BBH}$). A simple ''$n\sigma v$'' calculation leads to,
\begin{eqnarray}
\frac{\Gamma_{\rm TDE}}{\rm gal}&=&N_{\rm BBH}\nnsc\Sigma_{\rm TDE} v_{\rm disp}=4.8\times 10^{-7}\ {\rm yr}^{-1}\times\nonumber\\
&\times&\left(\frac{\hat{\Sigma}}{0.01}\right)\left(\frac{N_{\rm BBH}}{100}\right)\left(\frac{m_{\rm BH}}{30\msun}\right)^{2}\times\nonumber\\
&\times&\left(\frac{10^5\mpcub}{\mnsc}\right)\left(\frac{\rhonsc}{10^5\mpcub}\right)^{1/2}\,.
\end{eqnarray}
Here, we have normalized to the typical values of $\hat{\Sigma}$ found in the previous Section and used the mean density ($\rhonsc$) of the host NSC. 

In the previous equation, the number of BBH has been normalized considering a standard \citet{kro01} IMF. This is true for stellar SBHs only and assuming a single burst of star formation, while NSCs have more complex histories, with episodic star formation and accretion of star clusters that can lead to morphological and structural transformations \citep{anto2013}. Moreover, dynamics may change the number of BBHs over time, for instance ejecting or disrupting some of them. For IMBHs, the numbers are even more unconstrained. A detailed calculation of the TDE rate in NSCs could be obtained, for example, by means of detailed and large N-body simulations of NSCs, similarly to the ones performed for globular clusters in \citet{Kremer2020}. This is beyond the scope of the present paper and we leave this investigation to a future study.

Interestingly, this predicted rate is of the same order of other TDE rates predicted in the literature for a number of different SBH and IMBH systems. Mergers of stars with SBHs have been estimated to $10^{-7}-10^{-9}$ yr$^{-1}$ in star clusters \citep{Kremer2020,sams2019} and to $10^{-1}-10^{-4}$ yr$^{-1}$ in hierarchical triple systems \citep{fraglpk2019}. Finally, the rates for SMBHs and IMBHs are estimated to be $10^{-4}-10^{-5}$ yr$^{-1}$ \citep{vanvelzen2018} and $10^{-3}-10^{-5}$ yr$^{-1}$ \citep{fl2018,fragleiginkoc18}, respectively. 

\section{Electromagnetic signatures of tidal disruption events in binary black holes}
\label{sec:electr}

In this Section, we start with the discussion of a canonical TDE from an isolated BH, and we then extend the discussion to the case of a binary BH-BH.

A star of mass $m_*$ approaching a BH of mass $m_{\rm BH}$ will be torn apart by its tidal forces if the pericenter $R_p$ of its orbit passes within the tidal radius $R_{\rm T}$ of the BH. In the Newtonian approximation, which is reasonably accurate for $R_T>>R_g \equiv Gm_{\rm BH}/c^2$ ($R_g$ being the BH gravitational radius), this is given by Eq.~\ref{eq:Rt}. The strength of the encounter is measured by the extent to which the pericenter of the star penetrates inside the tidal radius of the BH, quantified by the penetration parameter, $\beta_{\rm p}={R_{\rm T}}/{R_{\rm p}}$\,. The detailed properties of a TDE depend on the magnitude of $\beta_{\rm p}$, varying from a mild disruption for $\beta_{\rm p}\approx 1$ to an increasingly more violent encounter (and fuller disruption of the star) as $\beta_{\rm p}$ increases. Since the different fluid elements comprising the disrupted star are disrupted at different distances from the BH, they will have a spread in binding energy. The disrupted material with positive binding energy is able to escape the gravitational pull of the BH, while the bound gas stream returns, moving on nearly geodesic orbits around the BH. 

The timescale  $t_0$  over which the bound debris return to the BH is approximately determined by the time that it takes for the most bound debris to return to the pericenter $R_p$ (with $R_p\sim R_t$), 
\begin{eqnarray}
t_0 &=& \frac{\pi R^3_T}{\sqrt{2Gm_{\rm {BH}} R_*^3}} \approx \nonumber\\                                   &\approx & 9\times 10^4{\rm s} \left(\frac{R_{\rm T}}{R_\odot}\right)^3\left(\frac{R_\odot}{R_*}\right)^{3/2}
\left(\frac{10^3M_\odot}{m_{\rm {BH}}}\right)^{1/2}=\nonumber \\
&=&9\times 10^4{\rm s} \left(\frac{R_*}{R_\odot}\right)^{3/2}
\left(\frac{m_*}{M_\odot}\right)^{-1} \left(\frac{m_{\rm BH}}{10^3M_\odot}\right)^{1/2}\,.  
\label{eq:t0}                                       
\end{eqnarray} 
Once the first debris reaches the BH, the rate of fallback for the remaining bound debris  will depend on their energy distribution with mass. For a flat distribution, which is found in objects which are  completely or nearly disrupted, the return rate follows the scaling (\citealt{Phinney1989}, based on the original argument by \citealt{Rees1988}),
\begin{equation}                    
\dot{M}_{\rm fb}\sim \frac{M_{\rm fb,0}}{t_0}\left(\frac{t}{t_0}\right)^{-5/3}\,,        
\label{mdot}                                   
\end{equation}
where
\begin{eqnarray}
\dot{M}_{\rm fb,0}&=&\frac{M_{\rm fb,0}}{t_0}\approx 10^{-5}\msun\ {\rm s}^{-1}\times\nonumber\\
&\times& \left(\frac{m_*}{1\msun}\right)^2\left(\frac{R_*}{\rsun}\right)^{-3/2}\left(\frac{m_{\rm BH}}{10\msun}\right)^{-1/2}
\end{eqnarray}
On the other hand, in the cases for which only partial disruption of the star is achieved, the fallback rate is found to be steeper than $t^{-5/3}$ (e.\ g. \citealt{Guillochon2013}).

\begin{figure} 
\centering
\includegraphics[scale=0.55]{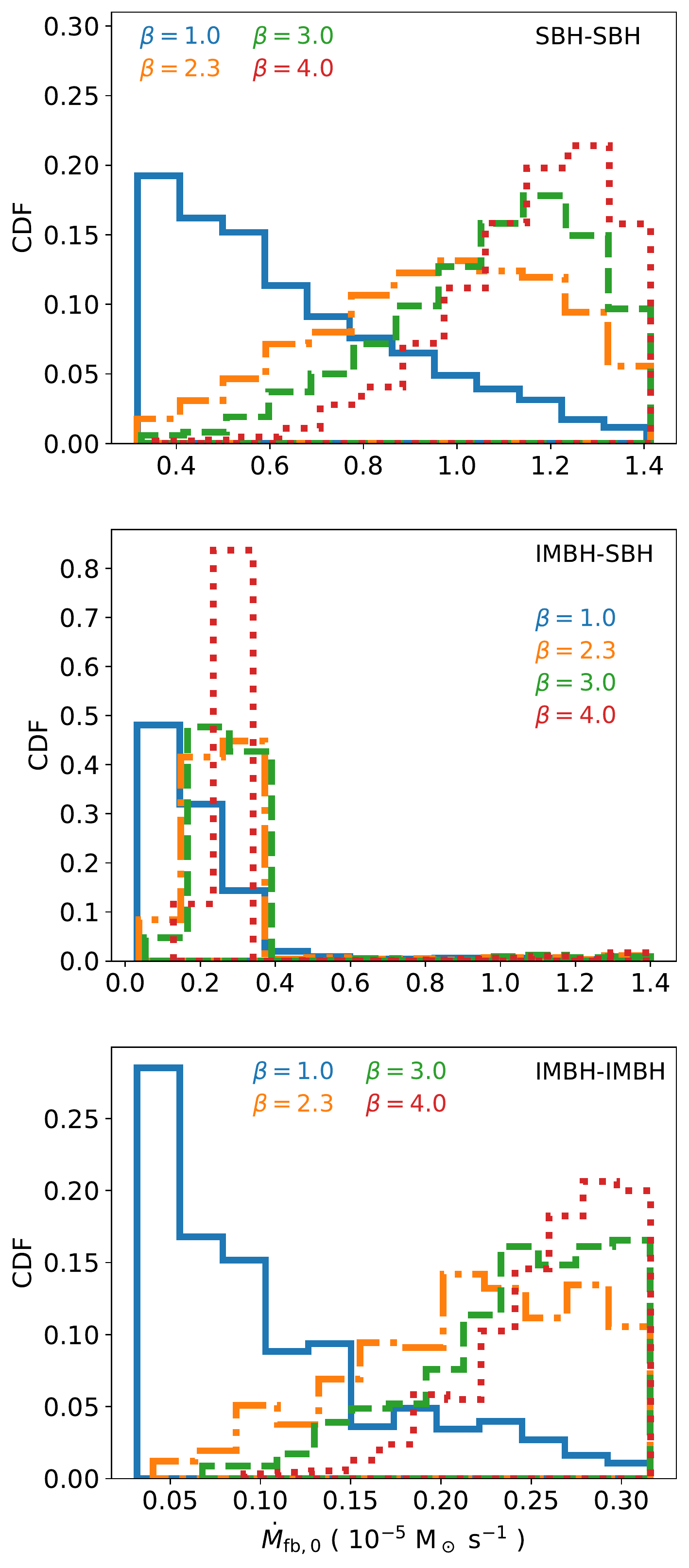}
\caption{Cumulative distribution functions of $\dot{M}_{\rm fb,0}$ for different values of the slope of the BH mass function, for SBH-SBH (top), IMBH-SBH (center), and IMBH-IMBH (bottom) binaries. The density and mass of the host NSC are assumed to be $\rhonsc=5\times 10^6\mpcub$ and $\mnsc=10^7\msun$, respectively.}
\label{fig:mdotall}
\end{figure}

Figure~\ref{fig:mdotall} shows the distribution of $\dot{M}_{\rm fb,0}$ for different values of the slope of the BH mass function, for SBH-SBH (top), IMBH-SBH (center), and IMBH-IMBH (bottom) binaries. The density and mass of the host NSC are $\rhonsc=5\times 10^6\mpcub$ and $\mnsc=10^7\msun$, respectively. Since $\dot{M}_{\rm fb,0}\propto m_{\rm BH}^{-1/2}$, steeper BH mass functions (higher $\beta$'s) produce larger $\dot{M}_{\rm fb,0}$'s. We find that the distribution of $\dot{M}_{\rm fb,0}$ peaks at $\sim 4\times 10^{-6}\msun$ s$^{-1}$ and $\sim 1.3\times 10^{-5}\msun$ s$^{-1}$ for $\beta=1$ and $\beta=4$, respectively, for an SBH-SBH binary. In the same way, the distribution of $\dot{M}_{\rm fb,0}$ presents a peak at $\sim 5\times 10^{-7}\msun$ s$^{-1}$ and $\sim 3\times 10^{-6}\msun$ s$^{-1}$ for $\beta=1$ and $\beta=4$, respectively, for an IMBH-SBH binary. The distributions for IMBH-SBH binaries present peaks similar to the distributions for IMBH-IMBH binaries, with tails up to $\sim 1.3\times 10^{-5}\msun$ s$^{-1}$. The reason is that most of the TDEs in these binaries are due to the disruption of stars by the more massive IMBH, which constitute the bulk of the distribution, while only a few TDEs are due to the SBH, which make up the tails of the distribution.

The detailed evolution of the falling-back bound material is largely determined by its ability to circularize in a disk. For this to happen, the bound matter must lose a significant amount of energy. This is believed to be possible as a result of a combination of effects, such as compression of the stream at pericenter, thermal viscous and magnetic shears, and the General Relativistic (GR) apsidal precession, which  forces highly eccentric debris streams to self-intersect (see i.e. \citealt{Chen2018} for a review of these effects). The issue of whether circularization can be completed before the end of the actual event still remains a matter of debate \citep{Piran2015}. The  energy dissipation due to GR apsidal crossings  increases with the BH mass \citep{Chen2018}, and can become sufficient to produce circularization. However, for IMBHs and SBHs,  this process  does not appear to provide enough energy dissipation to lead to circularization. On the other hand, from an observational point of view, the emission from a candidate TDE from an IMBH \citep{Lin2018} was found to display evidence for thermal emission, indicative of a  thermal thin disk, and hence of efficient circularization. For low-mass BHs, circularization may be aided by the fact that the bound debris are not highly eccentric \citep{Kremer2020}. Recent hydrodynamical simulations \citep{Lopez2019} have found that, for SBHs (with masses in the range of those measured via GWs), energy dissipation via shocks at pericenter can be significant enough to lead to efficient circularization. For simplicity, we continue our discussion assuming that a disk can be promptly formed, but keeping this important caveat in mind.

\begin{figure*} 
\centering
\includegraphics[scale=0.6]{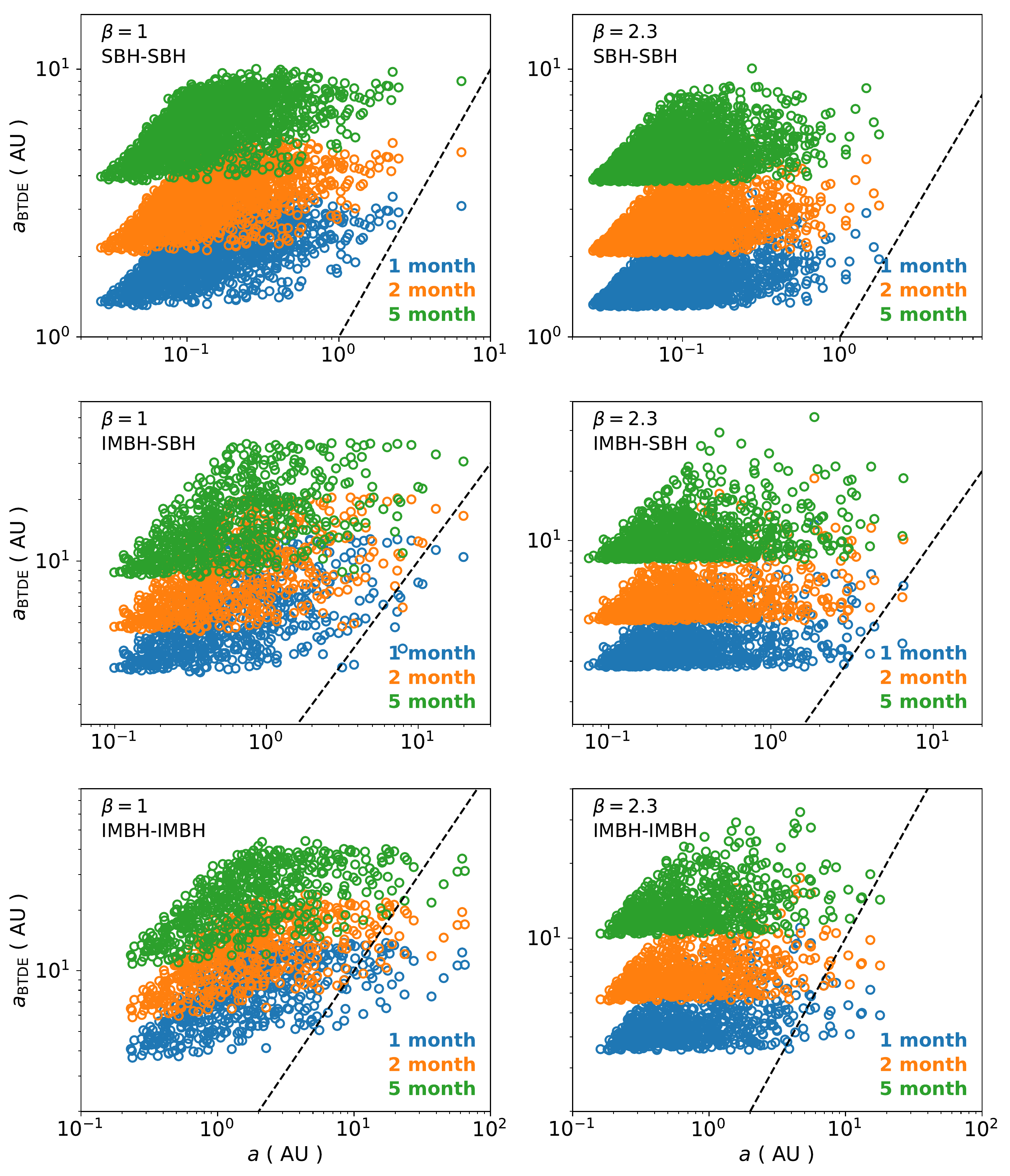}
\caption{Distribution of the orbital semi-major axis $a$ and $a_{\rm BTDE}$ (Eq.~\ref{eq:abtde}) for SBH-SBH (top), IMBH-SBH (center), IMBH-IMBH (bottom) binaries that undergo a TDE. Left: $\beta=1.0$; right $\beta=2.3$. The mass and density of the host NSC are assumed to be $\mnsc=10^6\msun$ and $\rhonsc=5\times 10^6\mpcub$, respectively. The black-dashed line represents the $x=y$ line. Different colors represent different observation times after the TDE: $1$ month (blue), $2$ month (orange), $5$ month (green).}
\label{fig:binarbeta}
\end{figure*}

\begin{figure*} 
\centering
\includegraphics[scale=0.6]{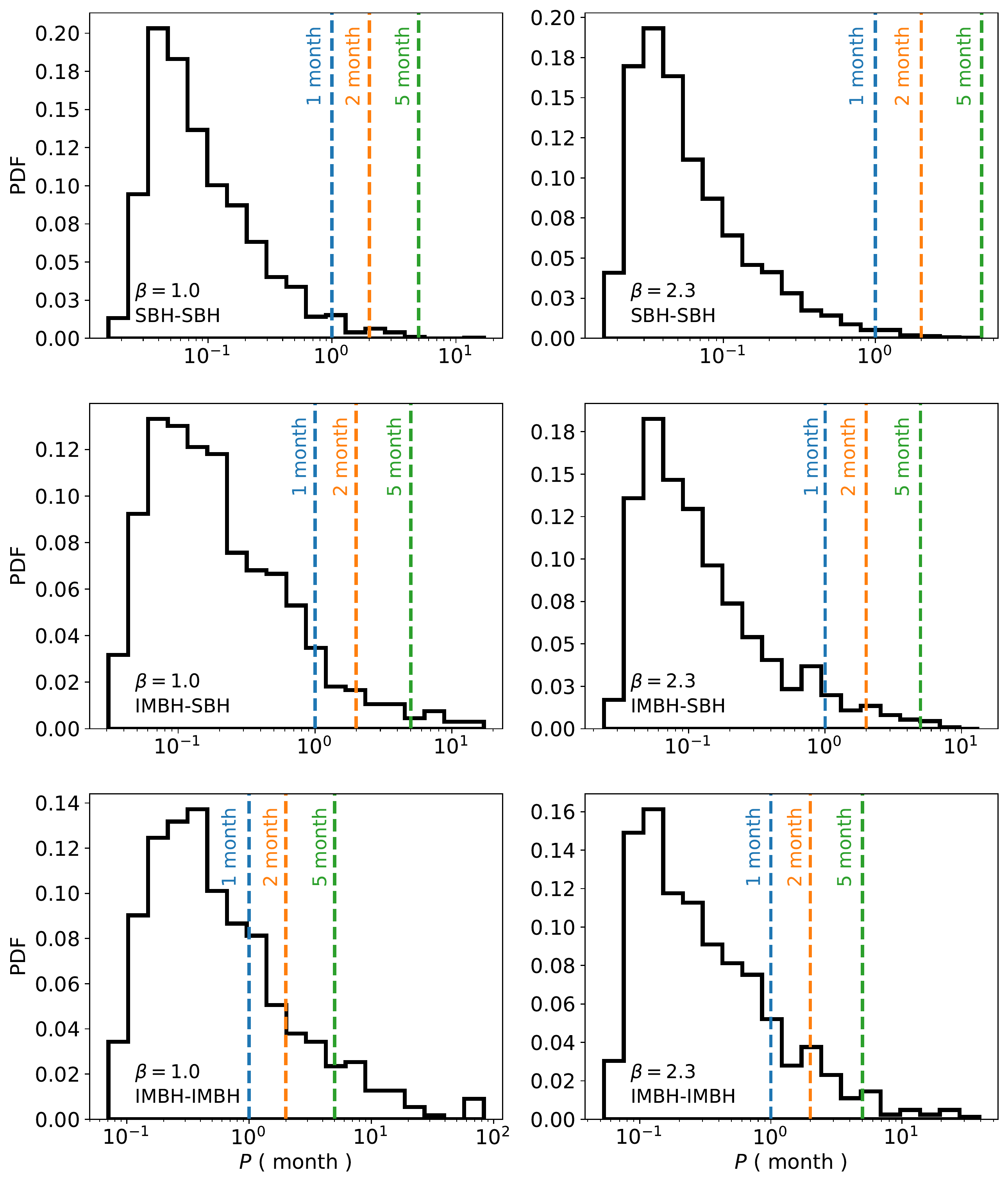}
\caption{Probability distribution function of the orbital periods $P$ for SBH-SBH (top), IMBH-SBH (center), IMBH-IMBH (bottom) binaries that undergo a TDE. Left: $\beta=1.0$; right $\beta=2.3$. The mass and density of the host NSC are $\mnsc=10^6\msun$ and $\rhonsc=5\times 10^6\mpcub$, respectively. The colored-dashed lines represent different observation times after the TDE: $1$ month (blue), $2$ month (orange), $5$ month (green).}
\label{fig:binarper}
\end{figure*}

Once a disk is formed, the timescale for the tidally disrupted debris to accrete is set by the viscous timescale \citep{Shakura1973}. For a geometrically
thick disk (i.e with a scale height $H/R\sim 1$), as expected at high accretion rates, this is given by,
\begin{equation}
t_{\rm acc}\sim \frac{1}{\alpha \Omega_{\rm K}(R_{\rm in})}=4\times 10^3\,{\rm s}\left(\frac{\alpha}{0.1}\right)^{-1}\left(\frac{R_{\rm in}}{2\rsun}\right)^{3/2}\left(\frac{m_{\rm BH}}{10^3\msun}\right)^{-1/2}\,.                      \label{eq:tvisc}
\end{equation}
If the viscous timescale is short compared to the fallback time, then the accretion rate to the BH directly tracks the fallback rate, i.e. $\dot{M}_{\rm acc}\sim \dot{M}_{\rm fb}$, otherwise its viscosity slows it down. However, even under these conditions, for typical parameters the early accretion rate is expected to be super-Eddington. Under these conditions, which are commonly observed in both the long and the short gamma-ray bursts (GRBs, e.g. \citealt{Popham1999,Narayan2001,Janiuk2004}), a relativistic jet could be launched, giving rise to high-energy emission. The maximum accretion rates for  SBHs have been estimated to be on the order of $10^{-6}-10^{-5}M_\odot$~s$^{-1}$. These values are somewhat comparable to those found in numerical simulations of the collapse of blue supergiant stars \citep{Perna2018}, which have been proposed as progenitors of the sub-class of very long GRBs \citep{Gendre2013}.  If launched \citep{fialk2017}, a relativistic jet would produce high energy radiation ($\gamma-$rays and X-rays) over timescales of $10^3-10^4$~s. However, even if a relativistic  jet is not launched, an outflow is likely to be driven, as demonstrated in both analytical works \citep{Narayan1994,Blandford1999} and in radiation-hydrodynamical simulations \citep{Dai2018,Kremer2020}. 

Since many TDEs are observed with significant (and often dominant) optical/UV emission, it is believed that this must be produced due to reprocessing of higher-energy radiation at some distance $R_{\rm ph}$ within the outflow (see e.g. \citealt{Mockler2019}).  To zeroth order, the spectrum is thermal, with an effective temperature at the photosphere $T_{\rm ph}\simeq [L/(4\pi R^2_{\rm ph}\sigma]^{1/4}$. For parameters typical of an IMBH, the spectrum peaks in the UV until the photosphere recedes back to the radius at which the outflow is launched, and later it peaks in the far UV and X-ray bands \citep{Chen2018}. The outflow luminosity during the super-Eddington phase is not well known. There have been suggestions \citep[e.g.][]{King2016,Lin2017} that, for $\dot{M}_{\rm acc}\gtrsim \dot{M}_{\rm Edd}$, this luminosity scales logarithmically with $\dot{M}_{\rm acc}$, while some numerical simulations of super-Eddington accretion disks indicate that, under the presence of strong magnetic fields ($\sim 10^3$~G), the wind luminosity can be substantially larger than the Eddington value \citep{Sadowski2016}. For an IMBH of $\sim 10^4\msun$, the super-Edddington phase lasts for over $\sim 10$ years \citep[e.g.,][]{Chen2018}, but it is shorter for smaller masses ($t(\dot{M})\propto m_{\rm BH}^{1/2}$ \citep[e.g.,][]{Mockler2019}).  The radiation-hydrodynamic calculations by \citet{Kremer2020} show that, for a representative BH of $20 M_\odot$, the outflow can give rise to optical transients $\sim 10^{42}-10^{44}$~erg~s$^{-1}$ (with the different values mostly dependent on the mass outflow rate), from timescales of a few hours to a few days, slowly declining down to luminosities on the order of a few $\times 10^{40}$ -- a few $\times 10^{42}$~erg~s$^{-1}$ over a timescale of about $100$ days. Once the accretion rate becomes sub-Eddington, the outflow is likely quenched, and the luminosity scales with the accretion rate, eventually becoming sub-dominant with respect to the luminosity of the  accretion disk, as $\dot{M}_{\rm acc}$ declines. 

All the above considerations strictly hold for the case of a TDE onto an isolated BH, or for the case in which, if the BH is part of a binary, the companion BH does not perturb the dynamics of the TDE stream \citep{hayas2016}. The required conditions for this to be the case,  and the modifications  in case of influence by the secondary, are discussed in the following \citep[see][for a review]{Coughlin2019}. The dynamics of the fallback debris will be drastically altered if the relative magnitudes of the apocenter distance $R_p$ and the binary separation $a$ are such that $R_p>a/2$. At time $T$ after the disruption, this translates into the condition,
\begin{equation}
a\lesssim a_{\rm BTDE}\approx 6 \left(\frac{M}{10^3M_\odot}  \right)^{1/3}\,\left(\frac{T}{\rm 1\,month}\right)^{2/3}\, {\rm AU}\,.
\label{eq:abtde}
\end{equation}
Given the typical tight separations of our BH-BH binaries, dictated by the high masses and densities of the host NSCs, it is evident that binarity will influence the TDEs discussed in this work.

We illustrate the effect of the secondary BH in Figure~\ref{fig:binarbeta}, where we plot the distribution of the orbital semi-major axis $a$ and $a_{\rm BTDE}$ for SBH-SBH (top), IMBH-SBH (center), IMBH-IMBH (bottom) binaries that undergo a TDE, in a NSC of mass and density of $\mnsc=10^6\msun$ and $\rhonsc=5\times 10^6\mpcub$, respectively. Since typically SBH-SBH binaries need to be tighter than IMBH-SBH and IMBH-IMBH binaries to be hard binaries in a NSC environment (Eq.~\ref{eqn:etahard}), the dynamics of the fallback debris is likely to be perturbed by the secondary in a larger number of cases. We find that most of the SBH-SBH binaries that undergo a TDE present this effect within $\sim 1$ month after the TDE. On the other hand, IMBH-SBH and IMBH-IMBH binaries can show signatures of binarity on longer timescales after the TDE, owing to their typical larger separations. We show in Figure~\ref{fig:binarper} the probability distribution function of the orbital periods $P$ for SBH-SBH (top), IMBH-SBH (center), IMBH-IMBH (bottom) binaries that undergo a TDE, presented in the Figure~\ref{fig:binarbeta}. Most of the systems have orbital periods smaller than a few months, implying that the modulation of the tidal debris due to the companion BH can be imprinted in the signal.

Simulations of SBHs (in the LIGO/Virgo range) by \citet{Lopez2019} have shown that, for $R_t>a$, the binary BH ends up embedded in a circumbinary disk which accretes on both BHs on the viscous timescale. This situation would hence likely lead to a transient similar to what discussed in the case of a single BH. If the two BHs have unequal mass, such as in the case of an IMBH-SBH binary, the accretion luminosity would be dominated by that of the larger BH. As discussed, a different outcome can arise if $R_t\lesssim a$. In this case the disruption is dominated by the primary BH (closer to the star), but the detailed later dynamics of the debris is influenced by the secondary BH by an amount which depends on the orbital separation. \citet{Lopez2019} quantified this via $a_{90}$, that is the semi-major axis of the material which contains $90$\% of the mass, as measured inwards from the most bound star material. If $a_{\rm 90}< R_{\rm L}$, where $R_{\rm L}$ is the Roche lobe radius of the disrupting BH \citep{Eggleton1983}, then most of the debris will accrete onto the primary BH, with only a small perturbation by the secondary one, and hence the TDE will be similar to that of the isolated BH discussed above.  On the other hand, if  $a_{\rm 90}\gtrsim R_{\rm L}$, the tidal debris will be affected by the presence of the secondary BH, and accrete onto it, giving rise to more complex light curves, whose detailed shape depends on several parameters, including the mass ratio of the BHs in the binary. However, peak accretion rates are still found to be of the same order of magnitude as for the isolated BH case, hence leading confidence to the observability of TDE events under a wide range of conditions. It should be noted, however, that extracting the masses of the BHs in the case in which both contribute to the TDE will be significantly more difficult with respect to the case in which the TDE event is dominated by a single BH \citep{Mockler2019}. 

The next decade is expected to be transformative for transient astronomy, and especially so for TDEs, with the upcoming optical surveys, and in particular the Zwicky Transient Facility (ZTF)\footnote{\url{https://www.ztf.caltech.edu/}} and LSST\footnote{\url{https://www.lsst.org/lsst/}}. An event with an optical luminosity of  $\sim 10^{42}$~erg~s$^{-1}$ can be detected by ZTF up to a distance of about 150 Mpc.  However, the major observational breakthrough for TDEs  is expected with LSST, which will be sensitive to an event of luminosity $\sim 10^{42}$~erg~s$^{-1}$ up to a distance of $\sim 0.5$~Gpc with 15~s of integration time in standard routine sky scans in the $r$ band (yielding a limiting magnitude of $r \sim 24.5$).  

\section{Conclusions}
\label{sec:conc}

We have discussed the electromagnetic signature of TDEs of stars through the assembly and merger of binary SBHs and IMBHs in NSCs. The lightcurve of these TDEs could be interrupted and modulated by the companion BH on the orbital period of the binary \citep{liuli2009,cough2017,fraglpk2019}, enabling to probe their orbital period distribution \citep{samsing2019}. We have explored in detail the dynamics that drive stars in NSCs to be tidally disrupted by SBH-SBH, IMBH-IMBH, and IMBH-IMBH binaries. We have shown that the orbital properties and the masses of the binaries that exhibit a TDE are set by the NSC mass and density and by the slope of the BH mass function.

For typical NSC properties, we have estimated a merger rate of $\sim 10^{-6}$--$10^{-7}\ {\rm yr}^{-1}$ per galaxy, which is similar to other TDE rates predicted in literature for a number of different SBH and IMBH system, including TDEs in star clusters \citep{Kremer2020,sams2019}, in hierarchical triple systems \citep{fraglpk2019}, and onto SMBHs in galactic nuclei \citep{vanvelzen2018} and IMBHs both in galactic nuclei and star clusters \citep{fl2018,fragleiginkoc18}.

 The ejected mass associated with these TDEs could produce optical transients of luminosity $\sim 10^{42}$--$10^{44}$ erg s$^{-1}$ with timescales of order a day to a month. These events should be detectable by optical transient surveys, such as ZTF and LSST. 

Present and upcoming GW detectors promise to detect hundreds of merging SBHs and IMBHs over the next decade. The origin and distribution of these SBH and IMBH binaries in dynamical environments is a fundamental and a key scientific question that will be addressed by the forthcoming data. TDEs offer a unique insight into dynamically assembled binaries in the dense environments of NSCs, where the effect of binarity can be detected within only a few weeks up to a few months. Moreover, they provide a unique probe of the elusive population of IMBHs, which could be difficult to trace through other observations. As new electromagnetic surveys, such as ZTF and LSST, improve and start operating, a large number of TDEs is expected to be observed, rendering these transients  excellent probes of the SBH and IMBH populations in galactic nuclei.\\
\ \\
\textit{Data Availability}\\
\ \\
The data underlying this article will be shared on reasonable request to the corresponding author.\\

\section*{Acknowledgements}

GF acknowledges support from a CIERA postdoctoral fellowship at Northwestern University. RP acknowledges support from NSF award AST-1616157. AL was supported in part by the Black Hole initiative at Harvard University, which is funded by JTF and GBMF grants.

\bibliographystyle{mn2e}
\bibliography{refs}

\begin{thebibliography}{}

\bibitem[\protect\citeauthoryear{{Amaro-Seoane}, {Gair}, {Freitag}, {Miller},
  {Mandel}, {Cutler} \& {Babak}}{{Amaro-Seoane} et~al.}{2007}]{seoane2007}
{Amaro-Seoane} P.,  {Gair} J.~R.,  {Freitag} M.,  {Miller} M.~C.,  {Mandel} I.,
   {Cutler} C.~J.,    {Babak} S.,  2007, Classical and Quantum Gravity, 24,
  R113

\bibitem[\protect\citeauthoryear{{Antognini} \& {Thompson}}{{Antognini} \&
  {Thompson}}{2016}]{antogn16}
{Antognini} J.~M.~O.,  {Thompson} T.~A.,  2016, \mnras, 456, 4219

\bibitem[\protect\citeauthoryear{{Antonini}}{{Antonini}}{2013}]{anto2013}
{Antonini} F.,  2013, \apj, 763, 62

\bibitem[\protect\citeauthoryear{{Antonini}, {Gieles} \&
  {Gualandris}}{{Antonini} et~al.}{2019}]{anto2019}
{Antonini} F.,  {Gieles} M.,    {Gualandris} A.,  2019, \mnras, 486, 5008

\bibitem[\protect\citeauthoryear{{Antonini} \& {Perets}}{{Antonini} \&
  {Perets}}{2012}]{antoper12}
{Antonini} F.,  {Perets} H.~B.,  2012, \apj, 757, 27

\bibitem[\protect\citeauthoryear{{Antonini} \& {Rasio}}{{Antonini} \&
  {Rasio}}{2016}]{antoras2016}
{Antonini} F.,  {Rasio} F.~A.,  2016, \apj, 831, 187

\bibitem[\protect\citeauthoryear{{Antonini}, {Toonen} \& {Hamers}}{{Antonini}
  et~al.}{2017}]{ant17}
{Antonini} F.,  {Toonen} S.,    {Hamers} A.~S.,  2017, \apj, 841, 77

\bibitem[\protect\citeauthoryear{{Arca-Sedda} \& {Gualandris}}{{Arca-Sedda} \&
  {Gualandris}}{2018}]{agu18}
{Arca-Sedda} M.,  {Gualandris} A.,  2018, \mnras, 477, 4423

\bibitem[\protect\citeauthoryear{{Askar}, {Szkudlarek}, {Gondek-Rosi\'{n}ska},
  {Giersz} \& {Bulik}}{{Askar} et~al.}{2017}]{askar17}
{Askar} A.,  {Szkudlarek} M.,  {Gondek-Rosi\'{n}ska} D.,  {Giersz} M.,
  {Bulik} T.,  2017, \mnras, 464, L36

\bibitem[\protect\citeauthoryear{{Banerjee}}{{Banerjee}}{2018}]{baner18}
{Banerjee} S.,  2018, \mnras, 473, 909

\bibitem[\protect\citeauthoryear{{Bartos}, {Kocsis}, {Haiman} \&
  {M\'{a}rka}}{{Bartos} et~al.}{2017}]{bart17}
{Bartos} I.,  {Kocsis} B.,  {Haiman} Z.,    {M\'{a}rka} S.,  2017, \apj, 835,
  165

\bibitem[\protect\citeauthoryear{{Belczynski}, {Holz}, {Bulik} \&
  {O'Shaughnessy}}{{Belczynski} et~al.}{2016}]{bel16b}
{Belczynski} K.,  {Holz} D.~E.,  {Bulik} T.,    {O'Shaughnessy} R.,  2016,
  \nat, 534, 512

\bibitem[\protect\citeauthoryear{{Bellovary}, {Brooks}, {Colpi}, {Eracleous},
  {Holley-Bockelmann}, {Hornschemeier}, {Mayer}, {Natarajan}, {Slutsky} \&
  {Tremmel}}{{Bellovary} et~al.}{2019}]{bello2019}
{Bellovary} J.,  {Brooks} A.,  {Colpi} M.,  {Eracleous} M.,
  {Holley-Bockelmann} K.,  {Hornschemeier} A.,  {Mayer} L.,  {Natarajan} P.,
  {Slutsky} J.,    {Tremmel} M.,  2019, BAAS, 51, 175

\bibitem[\protect\citeauthoryear{{Blandford} \& {Begelman}}{{Blandford} \&
  {Begelman}}{1999}]{Blandford1999}
{Blandford} R.~D.,  {Begelman} M.~C.,  1999, \mnras, 303, L1

\bibitem[\protect\citeauthoryear{{Bromm}}{{Bromm}}{2013}]{bromm2013}
{Bromm} V.,  2013, Reports on Progress in Physics, 76, 112901

\bibitem[\protect\citeauthoryear{{Bromm} \& {Larson}}{{Bromm} \&
  {Larson}}{2004}]{bromm2004}
{Bromm} V.,  {Larson} R.~B.,  2004, \araa, 42, 79

\bibitem[\protect\citeauthoryear{{Capuzzo-Dolcetta} \&
  {Miocchi}}{{Capuzzo-Dolcetta} \& {Miocchi}}{2008}]{capuzz2008}
{Capuzzo-Dolcetta} R.,  {Miocchi} P.,  2008, \mnras, 388, L69

\bibitem[\protect\citeauthoryear{{Capuzzo-Dolcetta} \& {Tosta e
  Melo}}{{Capuzzo-Dolcetta} \& {Tosta e Melo}}{2017}]{capuzzo2017}
{Capuzzo-Dolcetta} R.,  {Tosta e Melo} I.,  2017, \mnras, 472, 4013

\bibitem[\protect\citeauthoryear{{Chen} \& {Shen}}{{Chen} \&
  {Shen}}{2018}]{Chen2018}
{Chen} J.-H.,  {Shen} R.-F.,  2018, \apj, 867, 20

\bibitem[\protect\citeauthoryear{{Chen}, {Madau}, {Sesana} \& {Liu}}{{Chen}
  et~al.}{2009}]{chen2009}
{Chen} X.,  {Madau} P.,  {Sesana} A.,    {Liu} F.~K.,  2009, \apjl, 697, L149

\bibitem[\protect\citeauthoryear{{Chen}, {Sesana}, {Madau} \& {Liu}}{{Chen}
  et~al.}{2011}]{chen2011}
{Chen} X.,  {Sesana} A.,  {Madau} P.,    {Liu} F.~K.,  2011, \apj, 729, 13

\bibitem[\protect\citeauthoryear{{C{\^o}t{\'e}}, {Piatek}, {Ferrarese},
  {Jord{\'a}n}, {Merritt}, {Peng}, {Ha{\textcommabelow s}egan}, {Blakeslee},
  {Mei}, {West}, {Milosavljevi{\'c}} \& {Tonry}}{{C{\^o}t{\'e}}
  et~al.}{2006}]{cote2006}
{C{\^o}t{\'e}} P.,  {Piatek} S.,  {Ferrarese} L.,  {Jord{\'a}n} A.,  {Merritt}
  D.,  {Peng} E.~W.,  {Ha{\textcommabelow s}egan} M.,  {Blakeslee} J.~P.,
  {Mei} S.,  {West} M.~J.,  {Milosavljevi{\'c}} M.,    {Tonry} J.~L.,  2006,
  \apjs, 165, 57

\bibitem[\protect\citeauthoryear{{Coughlin}, {Armitage}, {Lodato} \&
  {Nixon}}{{Coughlin} et~al.}{2019}]{Coughlin2019}
{Coughlin} E.~R.,  {Armitage} P.~J.,  {Lodato} G.,    {Nixon} C.~J.,  2019,
  \ssr, 215, 45

\bibitem[\protect\citeauthoryear{{Coughlin}, {Armitage}, {Nixon} \&
  {Begelman}}{{Coughlin} et~al.}{2017}]{cough2017}
{Coughlin} E.~R.,  {Armitage} P.~J.,  {Nixon} C.,    {Begelman} M.~C.,  2017,
  \mnras, 465, 3840

\bibitem[\protect\citeauthoryear{{Dai}, {McKinney}, {Roth}, {Ramirez-Ruiz} \&
  {Miller}}{{Dai} et~al.}{2018}]{Dai2018}
{Dai} L.,  {McKinney} J.~C.,  {Roth} N.,  {Ramirez-Ruiz} E.,    {Miller} M.~C.,
   2018, \apjl, 859, L20

\bibitem[\protect\citeauthoryear{{Di Carlo}, {Giacobbo}, {Mapelli}, {Pasquato},
  {Spera}, {Wang} \& {Haardt}}{{Di Carlo} et~al.}{2019}]{DiCarlo2019}
{Di Carlo} U.~N.,  {Giacobbo} N.,  {Mapelli} M.,  {Pasquato} M.,  {Spera} M.,
  {Wang} L.,    {Haardt} F.,  2019, \mnras, 487, 2947

\bibitem[\protect\citeauthoryear{{Di Carlo}, {Mapelli}, {Giacobbo}, {Spera},
  {Bouffanais}, {Rastello}, {Santoliquido}, {Pasquato}, {Ballone}, {Trani},
  {Torniamenti} \& {Haardt}}{{Di Carlo} et~al.}{2020}]{DiCarlo2020}
{Di Carlo} U.~N.,  {Mapelli} M.,  {Giacobbo} N.,  {Spera} M.,  {Bouffanais} Y.,
   {Rastello} S.,  {Santoliquido} F.,  {Pasquato} M.,  {Ballone} A.~r.,
  {Trani} A.~A.,  {Torniamenti} S.,    {Haardt} F.,  2020, arXiv e-prints, p.
  arXiv:2004.09525

\bibitem[\protect\citeauthoryear{{Eggleton}}{{Eggleton}}{1983}]{Eggleton1983}
{Eggleton} P.~P.,  1983, \apj, 268, 368

\bibitem[\protect\citeauthoryear{{Fialkov} \& {Loeb}}{{Fialkov} \&
  {Loeb}}{2017}]{fialk2017}
{Fialkov} A.,  {Loeb} A.,  2017, \mnras, 471, 4286

\bibitem[\protect\citeauthoryear{{Fragione} \& {Bromberg}}{{Fragione} \&
  {Bromberg}}{2019}]{fragbrom2019}
{Fragione} G.,  {Bromberg} O.,  2019, \mnras, 488, 4370

\bibitem[\protect\citeauthoryear{{Fragione}, {Ginsburg} \& {Kocsis}}{{Fragione}
  et~al.}{2018}]{fraginkoc18}
{Fragione} G.,  {Ginsburg} I.,    {Kocsis} B.,  2018, \apj, 856, 92

\bibitem[\protect\citeauthoryear{{Fragione}, {Grishin}, {Leigh}, {Perets} \&
  {Perna}}{{Fragione} et~al.}{2019}]{fragrish2018}
{Fragione} G.,  {Grishin} E.,  {Leigh} N. W.~C.,  {Perets} H.~B.,    {Perna}
  R.,  2019, \mnras, 488, 47

\bibitem[\protect\citeauthoryear{{Fragione} \& {Kocsis}}{{Fragione} \&
  {Kocsis}}{2018a}]{frak18}
{Fragione} G.,  {Kocsis} B.,  2018a, Phys Rev Lett, 121, 161103

\bibitem[\protect\citeauthoryear{{Fragione} \& {Kocsis}}{{Fragione} \&
  {Kocsis}}{2018b}]{fragkoc2018}
{Fragione} G.,  {Kocsis} B.,  2018b, Phys Rev Lett, 121, 161103

\bibitem[\protect\citeauthoryear{{Fragione} \& {Kocsis}}{{Fragione} \&
  {Kocsis}}{2019}]{fragk2019}
{Fragione} G.,  {Kocsis} B.,  2019, \mnras, 486, 4781

\bibitem[\protect\citeauthoryear{{Fragione} \& {Leigh}}{{Fragione} \&
  {Leigh}}{2018a}]{frle2018}
{Fragione} G.,  {Leigh} N.,  2018a, \mnras, 480, 5160

\bibitem[\protect\citeauthoryear{{Fragione} \& {Leigh}}{{Fragione} \&
  {Leigh}}{2018b}]{fl2018}
{Fragione} G.,  {Leigh} N.,  2018b, \mnras, 479, 3181

\bibitem[\protect\citeauthoryear{{Fragione}, {Leigh}, {Ginsburg} \&
  {Kocsis}}{{Fragione} et~al.}{2018}]{fragleiginkoc18}
{Fragione} G.,  {Leigh} N. W.~C.,  {Ginsburg} I.,    {Kocsis} B.,  2018, \apj,
  867, 119

\bibitem[\protect\citeauthoryear{{Fragione}, {Leigh}, {Perna} \&
  {Kocsis}}{{Fragione} et~al.}{2019}]{fraglpk2019}
{Fragione} G.,  {Leigh} N. W.~C.,  {Perna} R.,    {Kocsis} B.,  2019, \mnras,
  489, 727

\bibitem[\protect\citeauthoryear{{Fragione}, {Loeb}, {Kremer} \&
  {Rasio}}{{Fragione} et~al.}{2020}]{frag2020}
{Fragione} G.,  {Loeb} A.,  {Kremer} K.,    {Rasio} F.~A.,  2020, arXiv
  e-prints, p. arXiv:2002.02975

\bibitem[\protect\citeauthoryear{{Fragione} \& {Silk}}{{Fragione} \&
  {Silk}}{2020}]{fragsilk2020}
{Fragione} G.,  {Silk} J.,  2020, arXiv e-prints, p. arXiv:2006.01867

\bibitem[\protect\citeauthoryear{{Fregeau}, {Cheung}, {Portegies Zwart} \&
  {Rasio}}{{Fregeau} et~al.}{2004}]{fregeau2004}
{Fregeau} J.~M.,  {Cheung} P.,  {Portegies Zwart} S.~F.,    {Rasio} F.~A.,
  2004, \mnras, 352, 1

\bibitem[\protect\citeauthoryear{{Fryer}, {Woosley} \& {Heger}}{{Fryer}
  et~al.}{2001}]{fryer2001}
{Fryer} C.~L.,  {Woosley} S.~E.,    {Heger} A.,  2001, \apj, 550, 372

\bibitem[\protect\citeauthoryear{{Gendre}, {Stratta}, {Atteia}, {Basa},
  {Bo{\"e}r}, {Coward}, {Cutini}, {D'Elia}, {Howell}, {Klotz} \&
  {Piro}}{{Gendre} et~al.}{2013}]{Gendre2013}
{Gendre} B.,  {Stratta} G.,  {Atteia} J.~L.,  {Basa} S.,  {Bo{\"e}r} M.,
  {Coward} D.~M.,  {Cutini} S.,  {D'Elia} V.,  {Howell} E.~J.,  {Klotz} A.,
  {Piro} L.,  2013, \apj, 766, 30

\bibitem[\protect\citeauthoryear{{Georgiev} \& {B{\"o}ker}}{{Georgiev} \&
  {B{\"o}ker}}{2014}]{geor2014}
{Georgiev} I.~Y.,  {B{\"o}ker} T.,  2014, \mnras, 441, 3570

\bibitem[\protect\citeauthoryear{{Georgiev}, {B{\"o}ker}, {Leigh},
  {L{\"u}tzgendorf} \& {Neumayer}}{{Georgiev} et~al.}{2016}]{georg2016}
{Georgiev} I.~Y.,  {B{\"o}ker} T.,  {Leigh} N.,  {L{\"u}tzgendorf} N.,
  {Neumayer} N.,  2016, \mnras, 457, 2122

\bibitem[\protect\citeauthoryear{{Georgiev}, {Hilker}, {Puzia}, {Goudfrooij} \&
  {Baumgardt}}{{Georgiev} et~al.}{2009}]{georg2009}
{Georgiev} I.~Y.,  {Hilker} M.,  {Puzia} T.~H.,  {Goudfrooij} P.,
  {Baumgardt} H.,  2009, \mnras, 396, 1075

\bibitem[\protect\citeauthoryear{{Giacobbo} \& {Mapelli}}{{Giacobbo} \&
  {Mapelli}}{2018}]{giac2018G}
{Giacobbo} N.,  {Mapelli} M.,  2018, \mnras, 480, 2011

\bibitem[\protect\citeauthoryear{{Giersz}, {Leigh}, {Hypki}, {L\"{u}tzgendorf}
  \& {Askar}}{{Giersz} et~al.}{2015}]{gie15}
{Giersz} M.,  {Leigh} N.~W.,  {Hypki} A.,  {L\"{u}tzgendorf} N.,    {Askar} A.,
   2015, \mnras, 454, 3150

\bibitem[\protect\citeauthoryear{{Gnedin}, {Ostriker} \& {Tremaine}}{{Gnedin}
  et~al.}{2014}]{gne14}
{Gnedin} O.~Y.,  {Ostriker} J.~P.,    {Tremaine} S.,  2014, \apj, 785, 71

\bibitem[\protect\citeauthoryear{{Grishin}, {Perets} \& {Fragione}}{{Grishin}
  et~al.}{2018}]{grish18}
{Grishin} E.,  {Perets} H.~B.,    {Fragione} G.,  2018, \mnras, 481, 4907

\bibitem[\protect\citeauthoryear{{Guillochon} \& {Ramirez-Ruiz}}{{Guillochon}
  \& {Ramirez-Ruiz}}{2013}]{Guillochon2013}
{Guillochon} J.,  {Ramirez-Ruiz} E.,  2013, \apj, 767, 25

\bibitem[\protect\citeauthoryear{{G{\"u}rkan}, {Freitag} \&
  {Rasio}}{{G{\"u}rkan} et~al.}{2004}]{gurk2004}
{G{\"u}rkan} M.~A.,  {Freitag} M.,    {Rasio} F.~A.,  2004, \apj, 604, 632

\bibitem[\protect\citeauthoryear{{G{\"u}rkan} \& {Rasio}}{{G{\"u}rkan} \&
  {Rasio}}{2005}]{gurk2005}
{G{\"u}rkan} M.~A.,  {Rasio} F.~A.,  2005, \apj, 628, 236

\bibitem[\protect\citeauthoryear{{Hamers} \& {Thompson}}{{Hamers} \&
  {Thompson}}{2019}]{hamers2019}
{Hamers} A.~S.,  {Thompson} T.~A.,  2019, \apj, 883, 23

\bibitem[\protect\citeauthoryear{{Hamilton} \& {Rafikov}}{{Hamilton} \&
  {Rafikov}}{2019}]{hamil2019}
{Hamilton} C.,  {Rafikov} R.~R.,  2019, arXiv e-prints, p. arXiv:1907.00994

\bibitem[\protect\citeauthoryear{{Hayasaki} \& {Loeb}}{{Hayasaki} \&
  {Loeb}}{2016}]{hayas2016}
{Hayasaki} K.,  {Loeb} A.,  2016, Scientific Reports, 6, 35629

\bibitem[\protect\citeauthoryear{Heggie}{Heggie}{1975}]{Heggie1975}
Heggie D.~C.,  1975, Mon.~Not.~R.~Astron.~Soc, 173, 729

\bibitem[\protect\citeauthoryear{{Janiuk}, {Perna}, {Di Matteo} \&
  {Czerny}}{{Janiuk} et~al.}{2004}]{Janiuk2004}
{Janiuk} A.,  {Perna} R.,  {Di Matteo} T.,    {Czerny} B.,  2004, \mnras, 355,
  950

\bibitem[\protect\citeauthoryear{{King} \& {Muldrew}}{{King} \&
  {Muldrew}}{2016}]{King2016}
{King} A.,  {Muldrew} S.~I.,  2016, \mnras, 455, 1211

\bibitem[\protect\citeauthoryear{{Kremer}, {Lu}, {Rodriguez}, {Lachat} \&
  {Rasio}}{{Kremer} et~al.}{2019}]{Kremer2020}
{Kremer} K.,  {Lu} W.,  {Rodriguez} C.~L.,  {Lachat} M.,    {Rasio} F.~A.,
  2019, \apj, 881, 75

\bibitem[\protect\citeauthoryear{{Kremer}, {Spera}, {Becker}, {Chatterjee}, {Di
  Carlo}, {Fragione}, {Rodriguez}, {Ye} \& {Rasio}}{{Kremer}
  et~al.}{2020}]{kretal2020}
{Kremer} K.,  {Spera} M.,  {Becker} D.,  {Chatterjee} S.,  {Di Carlo} U.~N.,
  {Fragione} G.,  {Rodriguez} C.~L.,  {Ye} C.~S.,    {Rasio} F.~A.,  2020,
  arXiv e-prints, p. arXiv:2006.10771

\bibitem[\protect\citeauthoryear{{Kremer}, {Ye}, {Rui}, {Weatherford},
  {Chatterjee}, {Fragione}, {Rodriguez}, {Spera} \& {Rasio}}{{Kremer}
  et~al.}{2020}]{kr2020ApJS}
{Kremer} K.,  {Ye} C.~S.,  {Rui} N.~Z.,  {Weatherford} N.~C.,  {Chatterjee} S.,
   {Fragione} G.,  {Rodriguez} C.~L.,  {Spera} M.,    {Rasio} F.~A.,  2020,
  \apjs, 247, 48

\bibitem[\protect\citeauthoryear{{Kroupa}}{{Kroupa}}{2001}]{kro01}
{Kroupa} P.,  2001, \mnras, 322, 231

\bibitem[\protect\citeauthoryear{{Kruckow}, {Tauris}, {Langer}, {Kramer} \&
  {Izzard}}{{Kruckow} et~al.}{2018}]{kruc2018}
{Kruckow} M.~U.,  {Tauris} T.~M.,  {Langer} N.,  {Kramer} M.,    {Izzard}
  R.~G.,  2018, \mnras, 481, 1908

\bibitem[\protect\citeauthoryear{{Lee}}{{Lee}}{1995}]{lee1995}
{Lee} H.~M.,  1995, \mnras, 272, 605

\bibitem[\protect\citeauthoryear{{LIGO/Virgo Scientific
  Collaboration}}{{LIGO/Virgo Scientific Collaboration}}{2019a}]{gwcat2019}
{LIGO/Virgo Scientific Collaboration} 2019a, Physical Review X, 9, 031040

\bibitem[\protect\citeauthoryear{{LIGO/Virgo Scientific
  Collaboration}}{{LIGO/Virgo Scientific Collaboration}}{2019b}]{ligov2019}
{LIGO/Virgo Scientific Collaboration} 2019b, \prd, 100, 064064

\bibitem[\protect\citeauthoryear{{LIGO/Virgo Scientific
  Collaboration}}{{LIGO/Virgo Scientific Collaboration}}{2020a}]{gwbhO3}
{LIGO/Virgo Scientific Collaboration} 2020a, arXiv e-prints, p.
  arXiv:2004.08342

\bibitem[\protect\citeauthoryear{{LIGO/Virgo Scientific
  Collaboration}}{{LIGO/Virgo Scientific Collaboration}}{2020b}]{gwnsO3}
{LIGO/Virgo Scientific Collaboration} 2020b, \apjl, 892, L3

\bibitem[\protect\citeauthoryear{{Lin}, {Guillochon}, {Komossa},
  {Ramirez-Ruiz}, {Irwin}, {Maksym}, {Grupe}, {Godet}, {Webb}, {Barret},
  {Zauderer}, {Duc}, {Carrasco} \& {Gwyn}}{{Lin} et~al.}{2017}]{Lin2017}
{Lin} D.,  {Guillochon} J.,  {Komossa} S.,  {Ramirez-Ruiz} E.,  {Irwin} J.~A.,
  {Maksym} W.~P.,  {Grupe} D.,  {Godet} O.,  {Webb} N.~A.,  {Barret} D.,
  {Zauderer} B.~A.,  {Duc} P.-A.,  {Carrasco} E.~R.,    {Gwyn} S. D.~J.,  2017,
  Nature Astronomy, 1, 0033

\bibitem[\protect\citeauthoryear{{Lin}, {Strader}, {Carrasco}, {Page},
  {Romanowsky}, {Homan}, {Irwin}, {Remillard}, {Godet}, {Webb}, {Baumgardt},
  {Wijnands}, {Barret}, {Duc}, {Brodie} \& {Gwyn}}{{Lin}
  et~al.}{2018}]{Lin2018}
{Lin} D.,  {Strader} J.,  {Carrasco} E.~R.,  {Page} D.,  {Romanowsky} A.~J.,
  {Homan} J.,  {Irwin} J.~A.,  {Remillard} R.~A.,  {Godet} O.,  {Webb} N.~A.,
  {Baumgardt} H.,  {Wijnands} R.,  {Barret} D.,  {Duc} P.-A.,  {Brodie} J.~P.,
    {Gwyn} S. D.~J.,  2018, Nature Astronomy, 2, 656

\bibitem[\protect\citeauthoryear{{Liu} \& {Lai}}{{Liu} \&
  {Lai}}{2019}]{liu2019}
{Liu} B.,  {Lai} D.,  2019, \mnras, 483, 4060

\bibitem[\protect\citeauthoryear{{Liu}, {Li} \& {Chen}}{{Liu}
  et~al.}{2009}]{liuli2009}
{Liu} F.~K.,  {Li} S.,    {Chen} X.,  2009, \apjl, 706, L133

\bibitem[\protect\citeauthoryear{{Loeb} \& {Furlanetto}}{{Loeb} \&
  {Furlanetto}}{2013}]{loebfur2013}
{Loeb} A.,  {Furlanetto} S.~R.,  2013, {The First Galaxies in the Universe}

\bibitem[\protect\citeauthoryear{{Lopez} Martin, {Batta}, {Ramirez-Ruiz},
  {Martinez} \& {Samsing}}{{Lopez} et~al.}{2019}]{Lopez2019}
{Lopez} Martin J.,  {Batta} A.,  {Ramirez-Ruiz} E.,  {Martinez} I.,
  {Samsing} J.,  2019, \apj, 877, 56

\bibitem[\protect\citeauthoryear{{Madau} \& {Rees}}{{Madau} \&
  {Rees}}{2001}]{madau2001}
{Madau} P.,  {Rees} M.~J.,  2001, \apjl, 551, L27

\bibitem[\protect\citeauthoryear{{Mandel}, {Brown}, {Gair} \&
  {Miller}}{{Mandel} et~al.}{2008}]{mandel2008}
{Mandel} I.,  {Brown} D.~A.,  {Gair} J.~R.,    {Miller} M.~C.,  2008, \apj,
  681, 1431

\bibitem[\protect\citeauthoryear{{Mastrobuono-Battisti}, {Perets} \&
  {Loeb}}{{Mastrobuono-Battisti} et~al.}{2014}]{mast14}
{Mastrobuono-Battisti} A.,  {Perets} H.~B.,    {Loeb} A.,  2014, \apj, 796, 40

\bibitem[\protect\citeauthoryear{{McKernan}, {Ford}, {Kocsis}, {Lyra} \&
  {Winter}}{{McKernan} et~al.}{2014}]{McKernan+2014}
{McKernan} B.,  {Ford} K.~E.~S.,  {Kocsis} B.,  {Lyra} W.,    {Winter} L.~M.,
  2014, \mnras, 441, 900

\bibitem[\protect\citeauthoryear{{McKernan}, {Ford}, {Lyra} \&
  {Perets}}{{McKernan} et~al.}{2012}]{McKernan+2012}
{McKernan} B.,  {Ford} K.~E.~S.,  {Lyra} W.,    {Perets} H.~B.,  2012, \mnras,
  425, 460

\bibitem[\protect\citeauthoryear{{McKernan}, {Ford}, {O'Shaughnessy} \&
  {Wysocki}}{{McKernan} et~al.}{2019}]{mcker2019}
{McKernan} B.,  {Ford} K.~E.~S.,  {O'Shaughnessy} R.,    {Wysocki} D.,  2019,
  arXiv e-prints, p. arXiv:1907.04356

\bibitem[\protect\citeauthoryear{{Miller}}{{Miller}}{2009}]{miller2009}
{Miller} M.~C.,  2009, Classical and Quantum Gravity, 26, 094031

\bibitem[\protect\citeauthoryear{{Miller} \& {Lauburg}}{{Miller} \&
  {Lauburg}}{2009}]{mill2009}
{Miller} M.~C.,  {Lauburg} V.~M.,  2009, \apj, 692, 917

\bibitem[\protect\citeauthoryear{{Mockler}, {Guillochon} \&
  {Ramirez-Ruiz}}{{Mockler} et~al.}{2019}]{Mockler2019}
{Mockler} B.,  {Guillochon} J.,    {Ramirez-Ruiz} E.,  2019, \apj, 872, 151

\bibitem[\protect\citeauthoryear{{Narayan}, {Piran} \& {Kumar}}{{Narayan}
  et~al.}{2001}]{Narayan2001}
{Narayan} R.,  {Piran} T.,    {Kumar} P.,  2001, \apj, 557, 949

\bibitem[\protect\citeauthoryear{{Narayan} \& {Yi}}{{Narayan} \&
  {Yi}}{1994}]{Narayan1994}
{Narayan} R.,  {Yi} I.,  1994, \apjl, 428, L13

\bibitem[\protect\citeauthoryear{{O'Leary}, {Kocsis} \& {Loeb}}{{O'Leary}
  et~al.}{2009}]{olea09}
{O'Leary} R.~M.,  {Kocsis} B.,    {Loeb} A.,  2009, \mnras, 395, 2127

\bibitem[\protect\citeauthoryear{{Pechetti}, {Seth}, {Neumayer}, {Georgiev},
  {Kacharov} \& {den Brok}}{{Pechetti} et~al.}{2020}]{pechetti2020}
{Pechetti} R.,  {Seth} A.,  {Neumayer} N.,  {Georgiev} I.,  {Kacharov} N.,
  {den Brok} M.,  2020, \apj, 900, 32

\bibitem[\protect\citeauthoryear{{Perna}, {Lazzati} \& {Cantiello}}{{Perna}
  et~al.}{2018}]{Perna2018}
{Perna} R.,  {Lazzati} D.,    {Cantiello} M.,  2018, \apj, 859, 48

\bibitem[\protect\citeauthoryear{{Perna}, {Wang}, {Farr}, {Leigh} \&
  {Cantiello}}{{Perna} et~al.}{2019}]{Perna2019}
{Perna} R.,  {Wang} Y.-H.,  {Farr} W.~M.,  {Leigh} N.,    {Cantiello} M.,
  2019, \apjl, 878, L1

\bibitem[\protect\citeauthoryear{{Phinney}}{{Phinney}}{1989}]{Phinney1989}
{Phinney} E.~S.,  1989, in {Morris} M.,  ed., The Center of the Galaxy Vol.~136
  of IAU Symposium, {Manifestations of a Massive Black Hole in the Galactic
  Center}.
p.~543

\bibitem[\protect\citeauthoryear{{Piran}, {Svirski}, {Krolik}, {Cheng} \&
  {Shiokawa}}{{Piran} et~al.}{2015}]{Piran2015}
{Piran} T.,  {Svirski} G.,  {Krolik} J.,  {Cheng} R.~M.,    {Shiokawa} H.,
  2015, \apj, 806, 164

\bibitem[\protect\citeauthoryear{{Popham}, {Woosley} \& {Fryer}}{{Popham}
  et~al.}{1999}]{Popham1999}
{Popham} R.,  {Woosley} S.~E.,    {Fryer} C.,  1999, \apj, 518, 356

\bibitem[\protect\citeauthoryear{{Portegies Zwart} \& {McMillan}}{{Portegies
  Zwart} \& {McMillan}}{2002}]{por02}
{Portegies Zwart} S.~F.,  {McMillan} S.~L.~W.,  2002, \apj, 576, 899

\bibitem[\protect\citeauthoryear{{Quinlan}}{{Quinlan}}{1996}]{quin1996}
{Quinlan} G.~D.,  1996, \na, 1, 35

\bibitem[\protect\citeauthoryear{{Rasskazov}, {Fragione} \&
  {Kocsis}}{{Rasskazov} et~al.}{2019}]{rassfk2019}
{Rasskazov} A.,  {Fragione} G.,    {Kocsis} B.,  2019, arXiv e-prints, p.
  arXiv:1912.07681

\bibitem[\protect\citeauthoryear{{Rasskazov}, {Fragione}, {Leigh}, {Tagawa},
  {Sesana}, {Price-Whelan} \& {Rossi}}{{Rasskazov} et~al.}{2019}]{rass2019}
{Rasskazov} A.,  {Fragione} G.,  {Leigh} N. W.~C.,  {Tagawa} H.,  {Sesana} A.,
  {Price-Whelan} A.,    {Rossi} E.~M.,  2019, \apj, 878, 17

\bibitem[\protect\citeauthoryear{{Rasskazov} \& {Kocsis}}{{Rasskazov} \&
  {Kocsis}}{2019}]{rasskoc2019}
{Rasskazov} A.,  {Kocsis} B.,  2019, \apj, 881, 20

\bibitem[\protect\citeauthoryear{{Rees}}{{Rees}}{1988}]{Rees1988}
{Rees} M.~J.,  1988, \nat, 333, 523

\bibitem[\protect\citeauthoryear{{Rodriguez}, {Amaro-Seoane}, {Chatterjee} \&
  {Rasio}}{{Rodriguez} et~al.}{2018}]{rod18}
{Rodriguez} C.~L.,  {Amaro-Seoane} P.,  {Chatterjee} S.,    {Rasio} F.~A.,
  2018, PRL, 120, 151101

\bibitem[\protect\citeauthoryear{{Sadowski} \& {Narayan}}{{Sadowski} \&
  {Narayan}}{2016}]{Sadowski2016}
{Sadowski} A.,  {Narayan} R.,  2016, \mnras, 456, 3929

\bibitem[\protect\citeauthoryear{{Samsing}}{{Samsing}}{2018}]{sam2018}
{Samsing} J.,  2018, \prd, 97, 103014

\bibitem[\protect\citeauthoryear{{Samsing}, {D'Orazio}, {Kremer}, {Rodriguez}
  \& {Askar}}{{Samsing} et~al.}{2019}]{sams2019}
{Samsing} J.,  {D'Orazio} D.~J.,  {Kremer} K.,  {Rodriguez} C.~L.,    {Askar}
  A.,  2019, arXiv e-prints, p. arXiv:1907.11231

\bibitem[\protect\citeauthoryear{{Samsing}, {Venumadhav}, {Dai}, {Martinez},
  {Batta}, {Lopez}, {Ramirez-Ruiz} \& {Kremer}}{{Samsing}
  et~al.}{2019}]{samsing2019}
{Samsing} J.,  {Venumadhav} T.,  {Dai} L.,  {Martinez} I.,  {Batta} A.,
  {Lopez} M.,  {Ramirez-Ruiz} E.,    {Kremer} K.,  2019, \prd, 100, 043009

\bibitem[\protect\citeauthoryear{{Secunda} et~al.,}{{Secunda}
  et~al.}{2018}]{secunda18}
{Secunda} B.,  et~al., 2018, ArXiv e-prints

\bibitem[\protect\citeauthoryear{{Shakura} \& {Sunyaev}}{{Shakura} \&
  {Sunyaev}}{1973}]{Shakura1973}
{Shakura} N.~I.,  {Sunyaev} R.~A.,  1973, \aap, 24, 337

\bibitem[\protect\citeauthoryear{{Silsbee} \& {Tremaine}}{{Silsbee} \&
  {Tremaine}}{2017}]{sil17}
{Silsbee} K.,  {Tremaine} S.,  2017, \apj, 836, 39

\bibitem[\protect\citeauthoryear{{Tagawa}, {Haiman} \& {Kocsis}}{{Tagawa}
  et~al.}{2019}]{Tagawa2019}
{Tagawa} H.,  {Haiman} Z.,    {Kocsis} B.,  2019, arXiv e-prints, p.
  arXiv:1909.10517

\bibitem[\protect\citeauthoryear{{Tremaine}, {Ostriker} \& {Spitzer}
  L.}{{Tremaine} et~al.}{1975}]{tremaine1975}
{Tremaine} S.~D.,  {Ostriker} J.~P.,    {Spitzer} L. J.,  1975, \apj, 196, 407

\bibitem[\protect\citeauthoryear{{Turner}}{{Turner}}{1977}]{Turner1977}
{Turner} M.,  1977, \apj, 216, 610

\bibitem[\protect\citeauthoryear{{Turner}, {C{\^o}t{\'e}}, {Ferrarese},
  {Jord{\'a}n}, {Blakeslee}, {Mei}, {Peng} \& {West}}{{Turner}
  et~al.}{2012}]{turner2012}
{Turner} M.~L.,  {C{\^o}t{\'e}} P.,  {Ferrarese} L.,  {Jord{\'a}n} A.,
  {Blakeslee} J.~P.,  {Mei} S.,  {Peng} E.~W.,    {West} M.~J.,  2012, \apjs,
  203, 5

\bibitem[\protect\citeauthoryear{{van Velzen}}{{van
  Velzen}}{2018}]{vanvelzen2018}
{van Velzen} S.,  2018, \apj, 852, 72

\bibitem[\protect\citeauthoryear{{Yang}, {Bartos}, {Gayathri}, {Ford},
  {Haiman}, {Klimenko}, {Kocsis}, {M{\'a}rka}, {M{\'a}rka}, {McKernan} \&
  {O'Shaughnessy}}{{Yang} et~al.}{2019}]{yang2019}
{Yang} Y.,  {Bartos} I.,  {Gayathri} V.,  {Ford} K.~E.~S.,  {Haiman} Z.,
  {Klimenko} S.,  {Kocsis} B.,  {M{\'a}rka} S.,  {M{\'a}rka} Z.,  {McKernan}
  B.,    {O'Shaughnessy} R.,  2019, Phys Rev Lett, 123, 181101

\end{thebibliography}

\end{document}